# Programing optical properties of single-walled carbon nanotubes with benzoyl peroxide derivatives of tailored chemical characteristics

Andrzej Dzienia [a, *], Patrycja Taborowska [a], Paweł Kubica-Cypek [a], Dawid Janas [a,*]

[a] Department of Chemistry, Silesian University of Technology, B. Krzywoustego 4, 44-100, Gliwice, Poland

* Corresponding authors: Andrzej.Dzienia@polsl.pl, Dawid.Janas@polsl.pl

## Abstract

Semiconducting single-walled carbon nanotubes (SWCNTs) have great potential for optoelectronics and photonics, further enhanced by covalent functionalization. However, scalable and controlled surface modification is challenging due to complex methodologies and unstable reagents. Benzoyl peroxide (BPO) has emerged as a simple alternative for introducing luminescent defects into SWCNTs. Yet, the lack of understanding of its radical chemistry limits precise defect engineering using BPOs. This is a major obstacle to the effective application of BPO in chemistry, despite its widespread use as a radical initiator. We present a thorough investigation into the radical chemistry of self-synthesized BPOs for functionalizing polymer-wrapped (6,5) and (7,5) SWCNTs in non-polar solvents, providing critical insights into the decomposition of BPO and its analogs. By varying the electronic and steric properties of typically unavailable BPO derivatives, we demonstrate tunability over the photoluminescence characteristics of SWCNTs, allowing control over defect density and light emission wavelength. This toolbox of BPO derivatives, created with simple radical chemistry and accessible organic precursors, alongside clarified structure-property relationships, facilitates effective implementation of BPO in chemical transformations and meticulous engineering of luminescent defects in SWCNTs for optoelectronic applications. Notably, this research offers insights into why SWCNTs modified with electron-deficient reactants provide the best optical characteristics.

# 1. Introduction

Semiconducting single-walled carbon nanotubes (*s*-SWCNTs) hold considerable promise for applications in optoelectronics, photonics, telecommunications, and quantum information technologies [1–3]. One of the key merits of SWCNTs is that their characteristics, including optical properties, are determined by their structure. Consequently, their optical absorption and emission wavelengths can be adjusted for a specific application. Taking into account that their capacity for structure-dependent light emission spans from the visible to the near-infrared range (NIR), SWCNTs are very valuable materials for basic and applied research. Unfortunately, a critical challenge in utilizing SWCNTs in optoelectronics and photonics is their relatively low photoluminescence quantum yield (PLQY), which greatly diminishes their implementation potential. Pristine SWCNTs typically exhibit poor PLQY (on the order of 0.5-2% [4,5]), meaning that the light they emit is unsatisfactorily dim.

Chemical functionalization of SWCNTs in water [6–8] or organic solvents [9–11] has emerged as a promising approach to overcome this limitation. Firstly, the inclusion of luminescent defects in SWCNTs, often referred to as Organic Color Centers (OCCs) or quantum defects [12], modifies the electronic structure of SWCNTs, traps the mobile excitons, and makes the radiative recombination of excitons more likely. Secondly, the formation of these exciton traps generates new emission peaks, such as $E_{11}^*$ and $E_{11}^{*-}$, which are redshifted relative to the native optical transition of SWCNTs ($E_{11}$). The extent of this shift, as well as the intensity of the newly formed peaks, depends on the nature of the attached functional groups [12] and their overall number on the SWCNT surface. For example, the $E_{11}^*$ peak is typically redshifted by 120–160 nm compared to $E_{11}$ [13], while further redshifted peaks, such as $E_{11}^{*-}$, can be induced by the attachment of divalent functional groups or through high degrees of functionalization [5,14]. Consequently, the PLQY of SWCNTs can be increased, thereby reaching 4% [5] or more. In light of the foregoing, it is evident that the development of effective functionalization strategies to improve the optical properties of SWCNTs is essential.

Diverse chemical modification methods enable the implantation of defects in SWCNTs, employing principles from both inorganic and organic chemistry, depending on the type of functional group introduced and the reagents utilized. The inorganic approach primarily focuses on oxidation, using reactive oxygen species. Effective oxidation methods include the use of hydrogen peroxide ($H_2O_2$), ozone ($O_3$), sodium hypochlorite (NaClO), unsaturated fatty acids, and organosulfur compounds [8]. The more complex organic strategy adapts a range of reactions typical of organic chemistry. Key strategies encompass diazonium coupling, employing diazonium salts ([$ArN_2$]X) [15], variants of reductive alkylation, such as Billups-Birch reduction (using lithium in liquid ammonia, $Li/NH_3$, followed by alkylation with RX) and alkylation using sodium naphthalenide (Na/naphthalene, then RBr). Alkylation methodologies also include dialkylation (e.g., using RLi followed by RX) [16]. Although these principal functionalization pathways – oxidation, alkylation, and arylation – provide the foundation for generating luminescent defects in SWCNTs that modulate their photoluminescence properties, they often lack the selectivity required for controlled defect implantation, limiting the potential for punctilious optical engineering of these nanomaterials.

To capitalize on the remarkable utility of benzoyl peroxide (BPO) in organic chemistry, we have recently explored how this compound can be used for covalent functionalization of SWCNTs [9,17]. The application of this reactant was found to facilitate the introduction of luminescent defects into SWCNTs [9], and the reaction conditions strongly influenced the resulting properties of BPO-modified SWCNTs [17]. Regrettably, despite the apparent structural simplicity of benzoyl peroxide (BPO), the underlying radical-driven chemical reactions exhibit considerable complexity [18–22] and remain poorly understood, particularly in the context of nanomaterial functionalization. A thorough comprehension of these mechanisms is crucial, given the wide range of applications of BPO across diverse fields of chemistry [23–25]. While the application of BPO in the functionalization of SWCNTs provides a relevant model system for analyzing these aspects, it should be acknowledged that only a narrow selection of BPO derivatives is commercially available. Having the ability to tune the chemical structure of BPO would enable gaining a more thorough understanding of the intricate nature of radical-based transformations. Concomitantly, such modifications would provide much more control over the physicochemical properties of BPO, including solubility, decomposition rate, chemical stability, and reactivity of the derived radicals [26–28] to identify better what are the application opportunities of this widely used chemical compound.

In response to these challenges, this study systematically explored the reactivity of a broad spectrum of self-synthesized BPO derivatives in the radical covalent functionalization of SWCNTs. By investigating how variations in the chemical structure of BPO derivatives (attachment of functional groups, e.g., alkyl, alkoxyl, carbonyl, halide, nitrile, or nitro in various positions) affect the radical generation and SWCNT functionalization, we deepened the understanding of the fundamental mechanisms governing these processes. Through careful control of reaction conditions, including temperature and radical type/concentration, we elucidated their roles in the course and the extent of decay of BPO derivatives. From the nanocarbon perspective, this knowledge opens up new prospects for fine-tuning the optical properties of SWCNT for making transformative applications such as sensors [29] or room-temperature single-photon emitters [30]. Furthermore, considering the vast potential of BPO-derived radicals for organic and polymer chemistry, the deduced relationships are highly valuable for designing more effective chemical transformations through radical chemistry. Last but not least, the results of this extensive synthetic initiative finally clarify why the electron-deficient reactants are most valuable for enriching the optical properties of SWCNTs.

## 2.    Experimental

A complete list of reagents, synthesis, and characterization procedures for polymers and peroxides used in this paper is provided in the Supporting Information (SI). It also contains a detailed description of other techniques used to carry out the investigation, in addition to those explained below, *i.e.*, Nuclear Magnetic Resonance ($^1$H NMR) and Size Exclusion Chromatography (SEC).

### 2.1. Preparation of (6,5)- and (7,5)-enriched SWCNTs for functionalization

The near-monochiral (6,5) or (7,5) SWCNT were collected according to the method presented earlier [31]. Briefly, in a typical suspension process, 6 mg of PFO-BPy (poly(9,9-dioctylfluorene-alt-6,6'-bipyridine)) or 9 mg of PFO (poly((9,9-dioctylfluorene)) synthesized in-house were dissolved in 5 mL of toluene. The solution was then transferred to a glass vial with 1.5 mg of pre-weighted SWCNTs. The mixture was homogenized in an ice-cooled bath sonicator to break the SWCNT bundles. Then, tip sonication was conducted while the mixture was cooled in an ice bath to keep a temperature of approximately 5 °C, which enabled the individualization of SWCNTs. After sonication, the thick suspension was transferred to a conical tube and centrifuged to remove the bundled SWCNTs and polymer aggregates. Then, 90% of the supernatant containing chirality-enriched (6,5) or (7,5) SWCNT fraction was transferred to a fresh vial for chemical modification and characterization.

### 2.2. Thermal functionalization of SWCNTs

The standard reaction consumed 1 mL of (6,5) or (7,5) SWCNTs diluted to 0.8 or 0.6 cm$^{-1}$, respectively ($E_{11}$ optical absorbance at the peak maximum). The molar concentrations of SWCNTs were calculated using molar absorptivity coefficients [32]. In parallel, the selected BPO derivative (aryl peroxide) was dissolved in toluene (1 mL) to achieve the necessary concentration ranging from 0.32 to 40 mg mL$^{-1}$. Then, the indicated components were combined in a glass vial and immersed in a stirred hot bath kept at 70 or 100°C. After an hour (unless indicated otherwise), the samples were removed, cooled down to room temperature, and analyzed.

### 2.3. Characterization of functionalized SWCNTs

Optical absorption spectra were recorded with a Hitachi U-2910 spectrophotometer using a quartz cuvette (5 mm path length) with a pure toluene-containing cuvette placed in the reference channel. The data were normalized to the peak of the $E_{11}$ transition whenever necessary for comparison of chirality distribution.

PL excitation-emission maps were collected with a ClaIR plate reader (Photonetc. Inc.) equipped with supercontinuum laser EVO HP EU-4 (NKT Photonics) and bandpass filter LLTF Contrast (NKT Photonics). The following settings were used for analysis (exposure time: 100 ms, excitation: 460-900 nm, emission: 947-1650 nm). For PL measurements, SWCNTs were diluted so that their concentration in the measured samples was low, *i.e.*, 0.27 µg mL$^{-1}$ (optical density of 0.15 in case of non-functionalized (6,5) SWCNTs) to avoid the inner-filter effect [33]. The PL spectra were extracted from the PL excitation-emission maps for the 574 nm excitation wavelength in case of (6,5) SWCNTs and 653 nm in case of (7,5) SWCNTs, normalized to the $E_{11}$ or $E_{11}*$ intensity for spectral shape comparison, and fitted with a set of Voigt functions using self-developed Python scripts (more details in SI). The area ratio of all defect-induced peaks to the $E_{11}$ peak was used for the estimation of the relative defect density implanted in the SWCNTs.

To obtain Raman spectra of the modified SWCNTs, it was necessary to prepare highly concentrated dispersions before drop-casting. The solvent excess was removed by centrifugation as follows. At least 1 mL of a sample containing the SWCNT suspension (with $E_{11}$ absorption in the range of 0.3-0.4 a.u. for pure SWCNTs) was mixed with 0.2 mL methanol (to promote precipitation of the SWCNTs) in a 2 mL conical tube and centrifuged. The supernatant generated from each sample was discarded after it was confirmed using PL excitation-emission mapping that it did not contain SWCNTs. The SWCNT material deposited in the conical tube was redispersed in 0.2 mL toluene by a 3-minute-long bath sonication and subsequently drop-casted on a glass substrate for characterization. The samples were characterized using a Renishaw inVia Raman microscope equipped with a 50× objective (Leica). The spectra were baseline-corrected and normalized to the maximum value of G peak intensity.

## 3. Results and discussion

While the commercial availability of BPO derivatives is somewhat limited, or the price thereof prohibitive in some instances, their synthesis at the laboratory scale presents a highly advantageous, reliable, and time-efficient alternative [23]. Adherence to specific safety protocols is paramount when handling peroxides to minimize the risk of premature decomposition. This risk can arise from contact with incompatible materials, such as acids, bases, metals, and certain polar organic solvents. It is recommended to limit reaction scales to 1-2 g. Furthermore, optimal storage conditions are essential, including the utilization of appropriately sized plastic containers and the avoidance of exposure to heat or radiation. The systematic application of these guidelines facilitates tonnage-scale production and utilization of BPO in industrial applications (as an active pharmaceutical ingredient, food additive, or radical initiator). At the same time, at the laboratory scale, it mitigates the risk of uncontrolled decay. The economic viability and ready availability of various carboxylic acids or their acid chlorides, commonly used as peroxide precursors, further enhance the attractiveness of this synthetic route (Figure 1a).

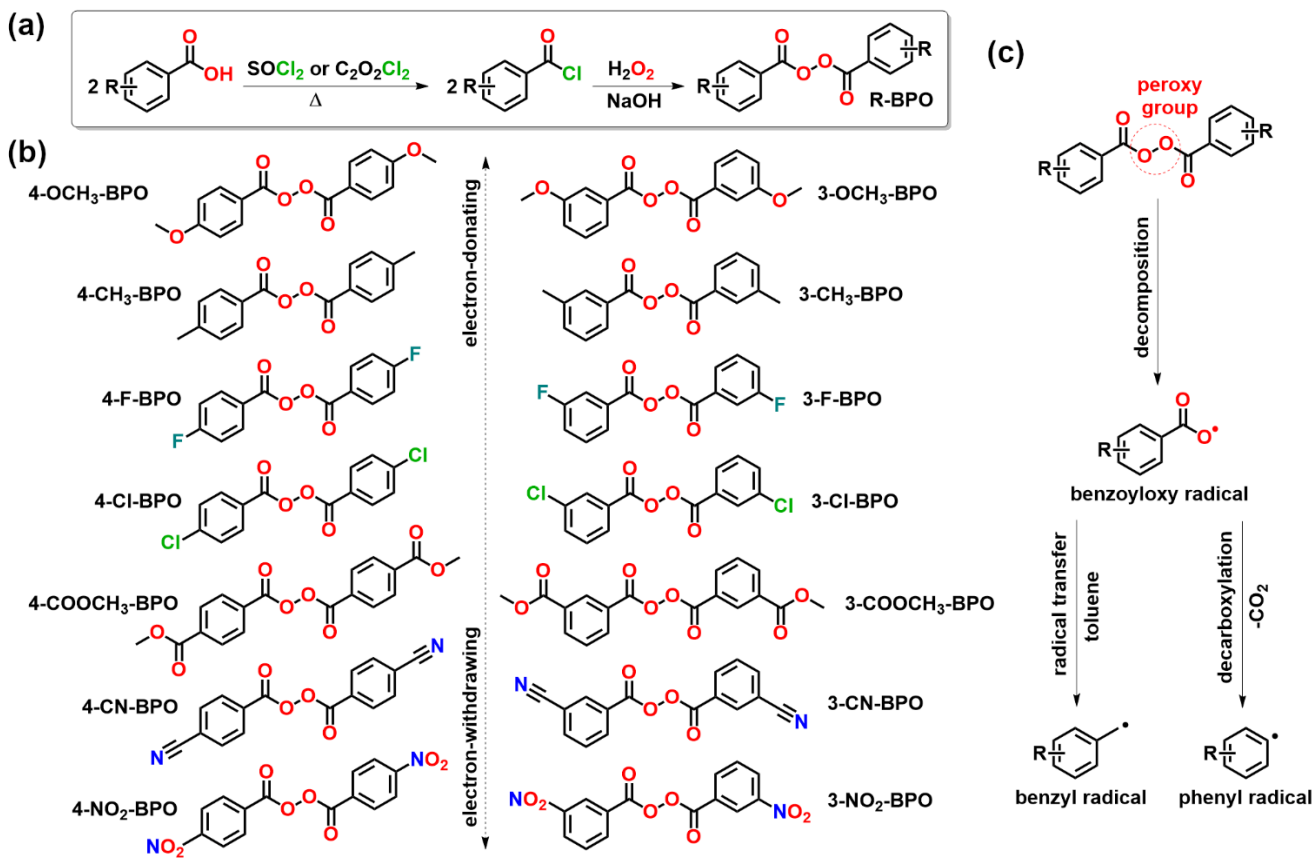

**Figure 1** a) Synthetic approach used to obtain BPO derivatives, b) a spectrum of BPO derivatives produced for SWCNT functionalization, c) decomposition of BPO derivatives giving rise to the generation of radicals for SWCNT functionalization.

The first step of the process is halogenation, which involves dissolving the carboxylic acid in a suitable solvent and stirring for just 1-4 hours at room temperature (or up to 40°C) in the presence of a halogen donor such as thionyl chloride or oxalyl chloride. This is followed by the removal of volatile byproducts and excess reagents via simple evaporation. Then, the acyl chloride coupling process is performed with 30% $H_2O_2$ and NaOH solution, which is also not time-consuming (1-2 hours). Moreover, the advantage is that the high purity of the resulting products (90-99%) minimizes the need for complex purification procedures (via crystallization or column chromatography), simplifying the work-up to straightforward steps of extraction, washing, and drying of the product. This approach enabled us to obtain a library of BPO derivatives (Figure 1b and Table S1) within a short timeframe with high yields and low material costs, which possessed unique and tailored properties (for details, see Section 1.4 in the SI).

Then, the chirality-enriched (6,5) and (7,5) SWCNT dispersions were obtained in large amounts in toluene using our recently published multi-cycle CPE procedure with in-house synthesized PFO-BPy or PFO [31]. High structural order and chiral purity of the resulting materials were confirmed by Raman (Figure S3a) and photoluminescence spectroscopy (Figure S3b), making them suitable for studying how the addition of luminescent defects impacts the optical properties. Specifically, Raman spectroscopy showed low $I_D/I_G$ band intensity ratios of ca. 0.02 for both SWCNT types, and PL excitation-emission analysis revealed a relatively small intensity of the photoluminescence sidebands (PSB) [34] both indicating minimal defect concentration and supporting the materials' high quality. The absorption spectra registered in the UV-Vis-NIR range showed a predominance of the (6,5) and (7,5) chiralities, the $E_{11}$ and $E_{22}$ optical transition bands of which were recorded (Figure S3c). Furthermore, UV-Vis-NIR spectroscopy revealed only trace amounts of other chiral species as minor components. Consequently, the overall purity of the starting materials was estimated to be approximately 85-90%, making them well-suited for investigating the influence of intentionally introduced luminescent defects on the materials' optical properties. Having characterized the pristine SWCNT dispersions, we proceeded with the covalent functionalization of the SWCNTs using various self-synthesized BPO derivatives according to the provided procedure. The purified dispersions of SWCNTs wrapped by PFO-BPy6,6' polymer in toluene were reacted with BPO derivatives at 70 or 100°C for 0.5–3 hours to create luminescent $sp^3$ defects. The molar ratios of the aryl peroxide to the SWCNTs ([R–BPO]/[CNT]) were in the range from 3:1 to 335:1 (details in Experimental Section).

As we showed earlier, the primary species involved in this process are benzoyloxyl radicals, while benzyl and phenyl radicals (generated in side reaction with toluene/decarboxylation of benzoyloxyl radicals, respectively) contribute to a lesser extent (Figure 1c) [17]. The increase in density of benzoyloxy groups attached to the SWCNT surface generally translates to increased intensity of the defect peak located at ca. 1160 nm wavelength in the PL spectrum. Moreover, the above-mentioned secondary radicals, also capable of attacking the surface of SWCNTs and changing their optical properties, should be considered. In light of this spectral complexity, deconvolution of the defect $E_{11}*$ peak was performed for selected samples to estimate the number and relative abundance of individual defect types (Figure S4). For ease of reference in the present work, defects are generally referred to as $E_{11}*$ or, whenever possible, more precisely as $E_{11}*_{(1160)}$, $E_{11}*_{(1220)}$, etc.

To demonstrate the differences in reactivity between BPO derivatives for (6,5) SWCNT functionalization, we present results obtained under different peroxide concentrations after 1-hour reactions at 70°C (Figure S5) and 100°C (Figure S7) analogously to our former publication wherein only non-tailored BPO was used [17]. For better comparison of the spectral shapes of the obtained defect peaks, we also provide the same spectra normalized to $E_{11}*$ in Figures S6 and S8, respectively. To systematically explore the behavior of various substituted BPOs, we analyze them one by one in the following subsections, taking into account their electronic character (electron-donating vs. electron-withdrawing) (Table S2). The influence of this rather unrecognized factor on SWCNT functionalization efficiency and PL response is discussed thoroughly below.

### 3.1. Carbon-based BPO substituents (4-CH$_3$, 3-CH$_3$)

The introduction of methyl substituents into the BPO structure provides an opportunity to assess how minor steric and electronic modifications affect the radical generation from variously modified BPOs and SWCNT functionalization efficiency. Due to its non-polar and weakly electron-donating (activating) nature, the methyl (-CH$_3$) group is expected to have minimal direct influence on radical stability but may affect decomposition pathways, e.g., via chain transfer or the decarboxylation process (Figure 1c), or alter steric accessibility of the reactive center in the case of the 3- substituent. At 70°C, neither BPO nor its methyl derivatives facilitated high-density defect implantation, as evidenced by the low intensity of the E$_{11}$* features in the PL spectra (Figure 2a-c). It should be noted that the functionalization process with BPO or 3-CH$_3$-BPO required an extremely high molar excess, i.e., ([R-BPO]/[CNT] = 168:1 or 84:1), to reach an evident increase in the intensity of the E$_{11}$* peak. Importantly, no functionalization was observed for 4-CH$_3$-BPO (Figure 2c) at this temperature, regardless of the concentration used. These findings suggested that at 70°C, either the radical formation process or the reactivity of the formed radicals was insufficient to drive covalent functionalization of SWCNTs.

To explain this observation, we compared the Hammett constants (summarized in Table S2) for BPO (-H) and its 4-CH$_3$ and 3-CH$_3$ derivatives. The values for BPO and 3-CH$_3$-BPO are similar ($\sigma = 0$ and $\sigma = -0.069$, respectively), whereas the 4-CH$_3$-BPO displays a lower Hammet constant value of $\sigma = -0.170$ (Figure 2d). Hence, the reactivity of the system was somehow related to the Hammet constants of the substituents. The 4-CH$_3$ ring-activating group hindered the possibility of SWCNT functionalization, confirming that the electron density distribution significantly influences radical reactivity and possibly its formation and decarboxylation rate [35]. This is reflected in the Atom-Centered Charges on the carbon atom adjacent to the carbonyl group (Table S2), indicated in red in Figure 2e. For the 4-CH$_3$-BPO reactant, the partial positive charge is slightly lower (0.062) than in the other cases (0.068 for 3-CH$_3$-BPO and 0.067 for unsubstituted BPO). This difference in electron density at the ortho and para positions relative to the CH$_3$ group is attributed to the resonance effects [36]. The lowest reactivity of the electron-donating 4-CH$_3$ substituent, which is supposed to *activate* the molecule, deserves clarification. Hammet substituent constants, whether ring-activating or deactivating, are considered from the perspective of a possible electrophilic attack. However, radicals can be either electrophilic or nucleophilic, depending on their structure, 37, which, as will be shown in this study, can significantly affect their capacity for SWCNT modification. Hence, the customarily defined activating nature of a given substituent does not necessarily mean that it will promote the functionalization of SWCNTs.

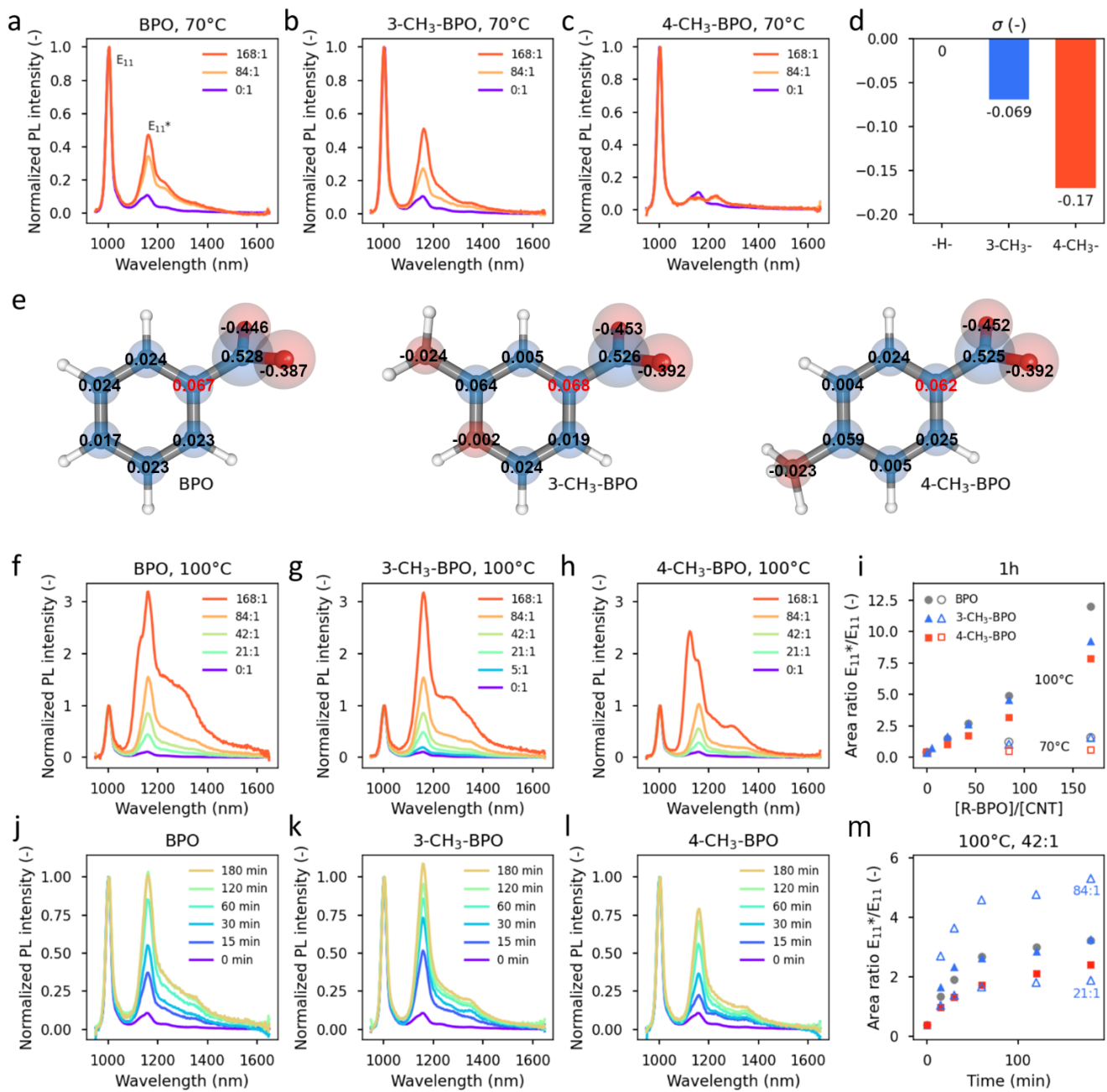

**Figure 2** PL spectra of SWCNTs functionalized using a) unsubstituted, or b) 3-CH₃- and c) 4-CH₃- substituted BPO in 1 hour-long reaction at 70 °C, using different molar concentrations of radical initiators [R-BPO]/[CNT] (in legends). d) Values of Hammet substituent constants. e) Calculated Atom-Centered Charges for radicals derived from unsubstituted BPO, 3-CH₃-BPO, and 4-CH₃-BPO, f-h) PL spectra analogous to the ones presented in a-c) but obtained at 100°C. i) The area ratios of $E_{11}^*/E_{11}$ peaks in spectra from a-c and f-h versus [R-BPO]/[CNT] concentration ratios. For better visibility, values obtained using 3- and 4-substituted BPOs were plotted using triangles and squares, respectively. j-l) PL spectra obtained for [R-BPO]/[CNT]=42:1, after a given time in minutes. m) Comparison of the increase in $E_{11}^*/E_{11}$ area ratios with reaction time for [R-BPO]/[CNT]=42:1 (full markers, the legend is the same as in i), but also 21:1 and 84:1 in case of 3-CH₃-BPO (empty markers).

When the reaction of SWCNTs with the same compounds was conducted at 100°C, distinct defect-related PL peaks emerged for all three reactants examined (Figure 2f-h), indicating the successful incorporation of a broad range of sp³-hybridized defects into the SWCNT lattice. The $E_{11}*/E_{11}$ area ratios (corresponding to the specific defect densities in SWCNTs [37]) were lower in the case of 4-CH₃-BPO processed at 100°C than the corresponding values for BPO and 3-CH₃-BPO (Figure 2i), confirming that the electron-donating properties of this substituent reduce electrophilicity of the resulting benzoyloxyl radical, making it less potent agent for SWCNT functionalization [27,38]. Up to the 84:1 molar ratio, the obtained PL spectra for both 3- and 4-CH₃-BPO closely resembled those obtained using non-substituted BPO. At a high radical excess of 168:1, SWCNT functionalization accelerated, judging by the elevated $E_{11}*/E_{11}$ ratios, and a new peak emerged at ca. 1130 nm, particularly for 4-CH₃-BPO (Figure 2h). This feature is consistent with previously reported signatures of benzyl radicals attached to SWCNTs [39]. Since this effect was also observed for the 3-OCH₃ substituent (in addition to 4-CH₃ and unsubstituted BPO), this phenomenon will be discussed collectively in the next section. At this point, we would like to stress that these additional defect peaks only emerge at very high covalent modification temperatures (100°C) and require substantial radical concentrations, which is not an issue in a typical approach.

A detailed analysis of the functionalization process under optimal conditions is shown in Figure 2j-m. Selective SWCNT functionalization was performed using lower concentrations of radical source ([R-BPO]/[CNT] = 42:1) to monitor this process in greater detail. The $E_{11}*/E_{11}$ ratios stabilized after 3 hours of reaction for all three reactants, suggesting that, at this point, the radicals were almost wholly consumed. As expected, the 4-CH₃-BPO, weakly activated by the presence of the methyl group in the para position, produced the lowest SWCNT defect densities. In contrast, the methyl group in the meta position did not seem to interfere with the reaction course, which will be explained in the subsequent parts of this article. At lower ratios of [R-BPO]/[CNT], adequately fewer defects were created, while at higher ratios, we again observed an accelerated reaction rate, manifested by a more rapid increase in the $E_{11}*/E_{11}$ area ratios, and overall, a greater defect density (Figure 2m).

### 3.2. Oxygen-containing BPO substituents (4-OCH₃, 3-OCH₃, 4-COOCH₃, 3-COOCH₃)

To understand the underlying relationships between the structure and reactivity of BPO derivatives more thoroughly, oxygen-containing substituents were engaged, which can affect the electronic characteristics of BPO derivatives to a greater extent [28]. The introduction of methoxy (-OCH₃) substituents into the BPO structure provided an opportunity to assess how stronger electron-donating functional groups influence radical formation and SWCNT functionalization. The Hammett constant indicated that 4-OCH₃ ($\sigma = -0.268$) was the most electron donating (ring-activating to electrophiles) substituent among those tested, whereas 3-OCH₃ ($\sigma = 0.115$) had a weakly electron withdrawing (ring-deactivating to electrophiles) nature. Despite this strong activating effect in electrophilic aromatic substitution reactions, the application of 4-OCH₃-BPO did not lead to the functionalization of SWCNTs under any tested conditions (Figures S5 and S7). This result stayed in accordance with earlier disclosed findings documenting that the weakly activated 4-CH₃-BPO compound was also not adequate for SWCNT modification. In contrast, placing the electron-rich methoxy group in the meta position (3-OCH₃-BPO) enabled significant functionalization at both high temperature (100°C) and lower temperature (70°C), provided that high radical concentrations were used. This effect can be explained by its weak electron-withdrawing characteristics with a much lower influence on the charge localized on a carbon atom adjacent to the carboxyl group (Figure S9). Consequently, compared to the pristine BPO, the reactivity of 3-OCH₃-BPO was considerably higher. Very similar defect densities to BPO in molar excess over SWCNTs of 21:1 were obtained using 11:1 excess of 3-OCH₃-BPO (Figure 3a). The optical spectra of the SWCNTs processed at both temperatures (70°C and 100°C) resembled that of the material treated with BPO (Figures S5 and S7).

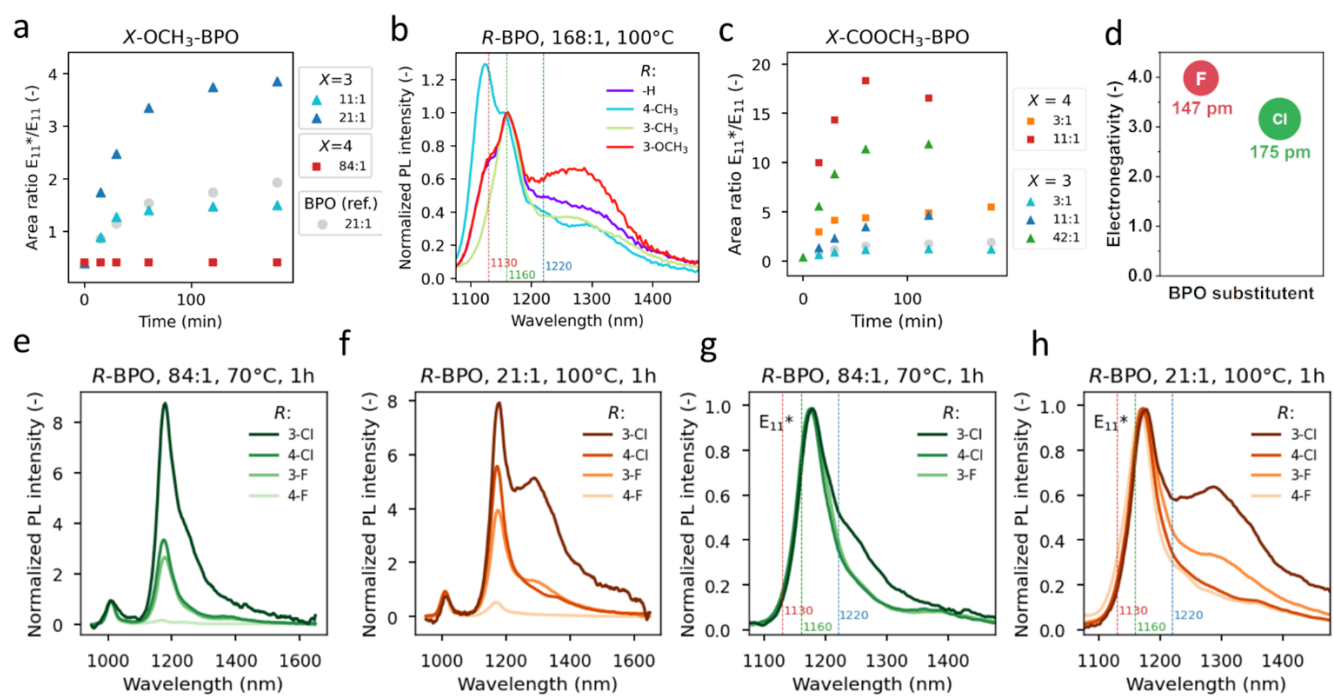

**Figure 3** Influence of oxygen-containing substituents on the course of SWCNT functionalization: a) Comparison of the increase in $E_{11}*/E_{11}$ area ratios with reaction time for 4- and 3-OCH₃-BPO compared to BPO. The reactions were conducted using different [R-BPO]/[CNT] concentration ratios listed in the legends. For better visibility, values obtained using 3- and 4-substituted BPOs were plotted using triangles and squares, respectively. b) Comparison of spectral shapes of the defect region $E_{11}*$, especially the $E_{11}*(1130)$ peak, obtained with large excess of different initiators. c) Comparison of the increase in $E_{11}*/E_{11}$ area ratios with reaction time for 4- and 3-COOCH₃-BPO as a function of BPO derivative concentration. Influence of halogen-derived substituents on the course of SWCNT functionalization: d) Comparison of electronegativity and van der Waals radii for F and Cl. Comparison of PL spectra obtained at e) lower temperature but higher [R-BPO]/[CNT] and f) higher temperature but lower [R-BPO]/[CNT] for -Cl and -F substituted BPO. g) and h) show spectral shapes of the $E_{11}*$ defect region in spectra from e) and f), respectively.

The spectrum of the material obtained at 100°C for a substantial initiator excess over the SWCNTs (168:1) also included a contribution from a peak at ca. 1130 nm, as well as another secondary peak at 1300 nm. It should be noted that the latter was present in significantly higher amounts for 3-OCH₃-BPO and appeared at a lower molar excess of the reactant to SWCNTs, proving higher reactivity of this BPO derivative. In our experiments, 4-CH₃-BPO exhibited the highest proportion of the peak at ca. 1130 nm, most likely corresponding to the benzyl modification of SWCNTs. This peak was also observed for unsubstituted BPO and 3-OCH₃-BPO, but under the investigated conditions, it did not appear for 3-CH₃-BPO (Figure 3b). How can we explain the increased radical concentration of the benzyl radicals in the examined samples, which may give rise to this curiosity? BPO, which generates aryloxyl radicals, is prone to decarboxylation [40,41]. In turn, the rate of decarboxylation of aryloxyl radicals, yielding secondary phenyl radicals, increases along the series 4-F-Ph-COO$^\bullet$ ≤ 4-OCH₃-Ph-COO$^\bullet$ < 4-CH₃-Ph-COO$^\bullet$ ≈ 4-Cl-Ph-COO$^\bullet$ < PhCOO$^\bullet$ < 3-Cl-Ph-COO$^\bullet$ [35]. In light of the preceding, we speculate that the reduced electrophilicity of the 4-CH₃-Ph-COO$^\bullet$ radical, due to its electron-rich nature, is responsible for its prolonged lifetime because it is less likely to functionalize SWCNTs. At the same time, this provides a greater opportunity for a radical transfer reaction with the solvent (toluene). As a result, phenyl radicals can be converted into more stable benzyl radicals via the cage effect – the phenomenon where surrounding solvent molecules temporarily confine reactive intermediates, influencing their reaction pathways – leading to the more stable species [22]. The importance of benzoyloxyl radical electrophilicity also seems supported by the lack of reactivity observed for 4-OCH₃-BPO, even at a high reactant-to-SWCNT ratio of 84:1 (Figure 3a). This compound is even more electron-rich due to the higher electron-donating capacity of the methoxy group positioned in the para position compared with the methyl moiety, which, as we presented above, did not favor SWCNT modification. Interestingly, prior research using diazonium salts successfully functionalized SWCNTs with the 4-OCH₃-Ph group [42], a result not achieved herein for the corresponding BPO derivative. This observation reinforces the distinct reactivity difference between benzoyloxyl (obtained by BPO decay) and phenyl radicals (generated differently from diazonium salts) toward the SWCNT surface, mirroring their varied affinity for the polymerization of unsaturated monomers [38]. Nonetheless, it should be noted that, in contrast to substituents in the 4-position, a substituent in the 3-position, such as a methyl or methoxy group, cannot engage in direct resonance stabilization with the radical center on the carboxyl group. Hence, despite the electron-donating nature of the methoxy group, 3-OCH₃-BPO functionalizes SWCNTs with a reactivity similar to that of unsubstituted BPO, unlike its highly deactivated 4-OCH₃ counterpart.

With this point clarified, let us turn to the analysis of ester substituents, which offer an insightful extension to the study of substituent effects. They are interesting because they are functional groups with both inductive and resonance effects. When the ester group is connected via its carbonyl carbon (*i.e.*, forming a Ph–COOCH₃ moiety), the group exerts a strong electron-withdrawing influence due to a combination of a −*I* (inductive) and a −*R* (resonance) effects. This is reflected in its Hammett constants ($\sigma = 0.37$ and $\sigma = 0.45$, for 3- and 4-positioned substituents, respectively). 4-COOCH$_3$–BPO proved effective in introducing defects into the SWCNT lattice at both 70°C and 100°C, in contrast to either methyl ($\sigma = -0.17$ and $\sigma = -0.069$, for 4- and 3- substituents, respectively) or methoxy-substituted analog ($\sigma = -0.268$ and $\sigma = 0.115$, for 4- and 3- substituents, respectively) as well as unsubstituted BPO ($\sigma = 0$). For the treatment of SWCNTs with 4-COOCH$_3$–BPO at 100°C, a defect density corresponding to an $E_{11}$*/$E_{11}$ area ratio of 3.4 was observed for one of the lowest radical concentrations ([R-BPO]/[CNT] = 3), indicating superior performance in covalent functionalization of SWCNTs. At the lower temperature of 70°C, radical surface modification was still efficient, necessitating a much lower excess of radical initiator over SWCNTs (11:1) to reach a significant $E_{11}$*/$E_{11}$ area ratio of 4.3, compared to other BPO-based compounds. These results suggest that the electron-withdrawing character of the 4-COOCH$_3$ positively influences the reactivity of the 4-substituted benzoyloxyl radicals during SWCNT functionalization. In a related observation, the 3-substituted derivative, 3-COOCH$_3$–BPO, exhibited lower functionalization efficiency at 100°C ($E_{11}$*/$E_{11}$ area ratio reaching 4.1 but for higher ratio [R-BPO]/[CNT] = 11) compared to its 4-isomer (Figure 3c). Moreover, 3-COOCH$_3$ substituted BPO was much less effective also at 70°C, even when a high molar excess of the initiator (84:1) was used. This behavior aligns with our previous findings, which indicated that both BPO derivatives modified in 3- and 4-positions lower the electron density of the aromatic ring. However, the 3-position is less electron-withdrawing because the resonance effect from the meta position cannot be effectively conjugated to the carbon atom bearing the peroxy group, unlike in the para isomer, and this difference results in the increased reactivity observed for 4-COOCH$_3$-BPO.

Analysis of optical spectra from 100°C (Figures S7 and S8) highlights additional striking differences in both reactivity and characteristics of resulted optical spectra between methoxy- and ester-substituted peroxides, as well as the influence of functional group position. Whereas 3-COOCH$_3$-BPO displayed a well-resolved additional peak at 1300 nm, 4-COOCH$_3$–BPO did not show significant spectral features beyond the primary $E_{11}$* defect registered at 1166 nm. Furthermore, the use of both 3-COOCH$_3$ and 3-OCH$_3$ for SWCNT functionalization led to a notable increase in the intensity of the peak at 1300 nm relative to optical spectra from unsubstituted BPO. We propose that this observation could be explained by the interaction of the substituent located in the 3-position with the SWCNT side wall. This interpretation stays in accordance with the findings of Yu et al., who recently showed that similar proximal modification of SWCNTs induces strongly redshifted PL peaks [43].

### 3.3. Halogen-containing BPO substituents (4-F, 3-F and 4-Cl, 3-Cl)

The introduction of halogen substituents into the BPO structure allows for a more in-depth evaluation of how electron-withdrawing groups with various atomic radii influence radical formation and (6,5) SWCNT functionalization efficiency. Fluorine and chlorine substituents are known to exert strong inductive effects while exhibiting limited resonance interactions. The Hammett constants indicate that 4-F ($\sigma = 0.062$) and 3-F ($\sigma = 0.337$) exert different electronic effects, with 3-fluorine being significantly more electron-withdrawing than its 4-counterpart. A similar trend is observed for chlorine, with 4-Cl ($\sigma = 0.227$) being less electron-withdrawing than 3-Cl ($\sigma = 0.373$).

At a lower temperature, both 3-Cl-BPO and 4-Cl-BPO, as well as 3-F-BPO, could introduce defects into the SWCNT lattice after 1 hour. When 4-F-BPO was used, no significant increase in $E_{11}*$ intensity was noticed, but prolonging the reaction time to 4 h revealed enhanced PL emission from the defect site. Even at 100°C, this initiator requires the highest [R-BPO]/[CNT] ratios of up to 168:1, whereas, for the remaining derivatives, the optimal ratio is 21:1 or 5:1 (or slightly higher to achieve $E_{11}*/E_{11}$ area ratio 4-5). By comparing the spectra of SWCNTs functionalized in identical conditions at both 70 or 100°C (Figure 3e and f, respectively), the order of reactivity with halogen-substituted peroxides was as follows: 4-F-BPO $\ll$ 3-F-BPO < 4-Cl-BPO $\ll$ 3-Cl-BPO. Due to the significant excess of the radical initiator over SWCNTs in these reactions, we observed differences in spectral shapes of the $E_{11}*$ defect region, derived from both 4- and 3-substituted fluorine and chlorine derivatives (Figure 3g, h), which can be attributed to their differing electronegativity and steric properties. The peak corresponding fluorescence at ca. 1130 nm was not observed in any case, whereas with increasing functional group size or for 3-position, the proportion of the band at ca. 1300 nm increased. Notably, for both 3-F and 3-Cl, this effect is apparent at much lower [R-BPO]/[CNT] molar ratios. At 100°C, it is observed for 21:1 (Figure 3h), in contrast to 4-Cl, where it is not observed until a reactant-to-SWCNT ratio of 84:1 is reached (Figure S8). Fluorine, being the most electronegative element, exerts a more substantial inductive effect (Figure 3d). However, its smaller atomic radius limits its electronic interactions with the SWCNT surface. Chlorine, while also electron-withdrawing, has a larger atomic radius, potentially altering steric interactions in the vicinity of the reaction site. These findings suggest that halogen substituents modulate radical reactivity through both electronic effects and steric hindrance. This underscores the importance of optimizing substituent placement, reaction mixture composition, and functionalization conditions when designing radical-based strategies for modifying SWCNT.

### 3.4. Nitrogen-containing BPO substituents (4-NO₂, 3-NO₂ and 4-CN, 3-CN)

Finally, the introduction of nitrogen-containing substituents, such as nitro (-NO₂) and cyano (-CN) groups, into the BPO structure provides insight into how strongly electron-withdrawing groups affect the radical formation process and SWCNT functionalization efficiency. Both substituents exhibit significant -*I* and -*R* effects, which influence the stability and reactivity of radical intermediates. The Hammett constants indicate that 4-NO₂ ($\sigma = 0.778$) and 4-CN ($\sigma = 0.830$) exert a strong electron-withdrawing influence, while their 3-counterparts, 3-NO$_2$ ($\sigma = 0.720$) and 3-CN ($\sigma = 0.560$), also significantly reduce electron density but to a lesser extent.

Functionalization experiments revealed that at 70°C, all nitrogen-containing substituents were effective, although 4-NO$_2$-BPO required not only high initiator concentrations to achieve functionalization but also prolonged reaction time. The E$_{11}$*/E$_{11}$ area ratio correlated well with radical initiator concentration (Figure S5). An exceptionally high peroxide concentration is required for 4-CN-BPO, which, together with 4-NO$_2$-BPO, exhibits low solubility in non-polar toluene. In contrast, significantly lower amounts are needed for 3-CN-BPO and 3-NO$_2$-BPO, up to 84:1 and 21:1, respectively. Interestingly, an enormous difference was noticed in the reactivity of NO$_2$-BPO, depending on the substituent position: higher defect densities were obtained with 3-NO$_2$-BPO at 70°C for 3:1 molar excess of the initiator over the SWCNTs than for 4-NO$_2$-BPO at 100°C for 168:1 ratio (Figure 4a). Among tested initiators, the one substituted with the nitro group in the 3-position was the most reactive, while the one with the same group in the 4-position was one of the least reactive compounds examined. For SWCNTs subjected to a high concentration of 3-NO$_2$-BPO, uncontrolled functionalization occurs, manifesting as a jagged optical spectrum and, over time, complete fluorescence quenching (Figure S5). This observation confirms published findings [44] that the process of SWCNT functionalization may be self-accelerating provided that reactant molecules are abundant, as the introduction of functional groups is more energetically favorable for covalently modified SWCNTs. As a result, multiple defects can accumulate in the affected region, ultimately compromising SWCNTs' optical properties.

At 100°C, all nitrogen-containing initiators are active, displaying radical reactivities analogous to those observed at lower temperatures, *i.e.*, 4-NO$_2$-BPO << 4-CN-BPO << 3-CN-BPO < 3-NO$_2$-BPO (Figure S7). Furthermore, the 3-substituted BPO derivatives require only small molar excesses, up to 11:1 and 3:1, respectively. This suggests that while both nitro and cyano groups withdraw electron density, positioning them at the 3-position, despite corresponding to lower Hammett constants, leads to more efficient covalent functionalization of SWCNTs. We hypothesize that this behavior stems from a combination of poor solubility of the 4-NO$_2$-BPO and 4-CN-BPO initiators in toluene, which limits their effective concentration and potentially the suboptimal reactivity profile of the derived radicals. While intense electron withdrawal of 4-isomers might be expected to increase reactivity, it could also lead to less selective radicals or faster decomposition pathways, whereas the 3-position offers a more effective balance between these processes. These findings highlight that although nitrogen-containing substituents are among the strongest electron-withdrawing groups tested, their influence on SWCNT functionalization strongly depends on their position within the BPO structure. The enhanced functionalization efficiency observed for 3-substituted derivatives suggests that fine-tuning electronic effects through substituent placement is crucial for optimizing defect formation in SWCNTs.

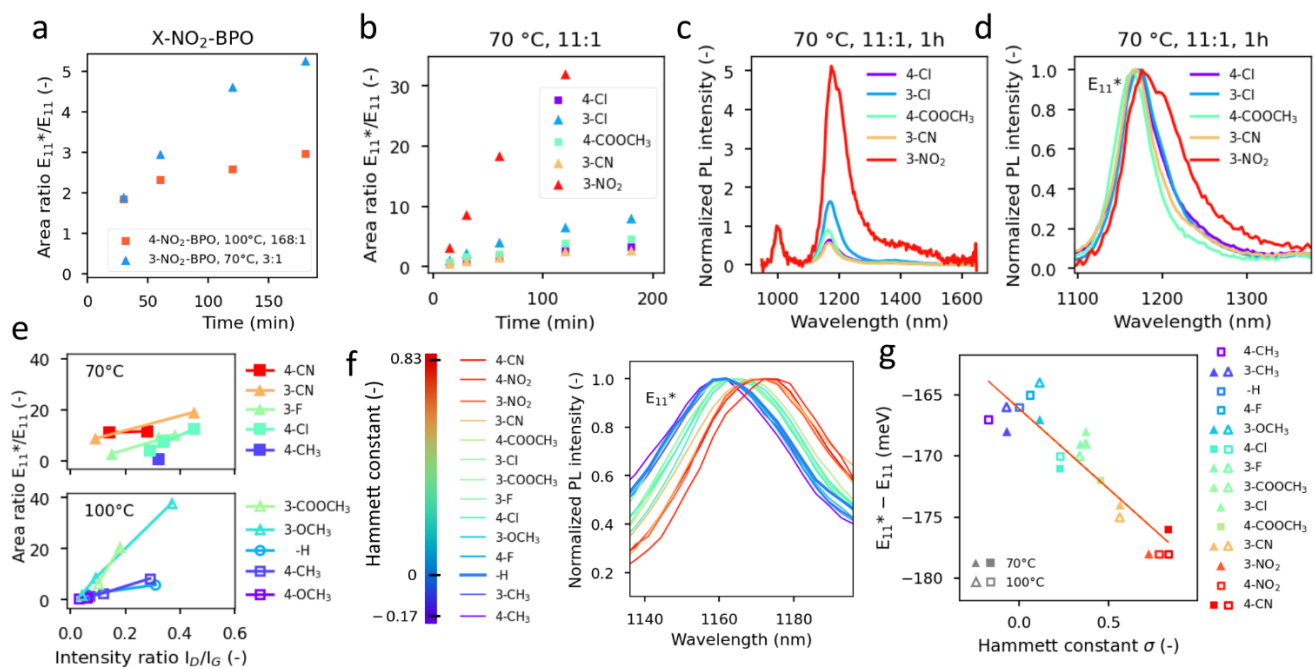

**Figure 4** a) Increase in area ratio of defect $E_{11}^*$ peak to original $E_{11}$ obtained with 3-NO$_2$-BPO, compared to 4-NO$_2$-BPO. b-d) Comparison of the reaction kinetics for the most reactive BPO initiators at the same temperature and with identical initiator-to-SWCNT molar ratios: b) increase in area ratio of $E_{11}^*$ peak to original $E_{11}$ peak; c and d) compared spectra obtained after 1 hour of reaction, normalized to $E_{11}$ and $E_{11}^*$ intensity, respectively. e) Relationship between the $E_{11}^*/E_{11}$ area ratio in PL spectra vs. intensity ratio of $I_D/I_G$ peaks in Raman spectra for (6,5) SWCNTs (data in Table S3). Colors and order in legends correspond to the values of the Hammett substituent constant. Lines are plotted only to improve visibility. f) Normalized $E_{11}^*$ peaks of (6,5) SWCNTs obtained upon functionalization using benzoyloxyl radicals with different substituents. Line colors were selected in linear dependence on the Hammet substituent constant, as shown in the colorbar. g) Close to the linear dependence of the experimentally observed optical trap depth on the Hammett constant. For better visibility, values obtained using 3- and 4-substituted BPOs were plotted using triangles and squares, respectively.

### 3.5. Direct comparison of most promising BPO derivatives and justification for their superior performance

We compared the functionalization kinetics, expressed as the $E_{11}^*/E_{11}$ area ratio, for the most reactive initiators under identical conditions (70°C and [R-BPO]/[CNT] = 11:1, Figure 4b-d). It confirmed our previous observations that the most reactive initiators are those with 3-NO$_2$ and 3-Cl groups, even though they differ significantly in their Hammett constants. Both the Cl and NO$_2$ groups in the 3- position exhibit weak resonance effects on the electron density of the benzoyloxyl radical, while both exert electron-withdrawing influence via induction. In radical chemistry, aside from radical reactivity, stabilization often plays a crucial role in determining the outcome. Although NO$_2$ exhibits a more substantial electron-withdrawing effect, the stabilization of the radical in the 3-position is already sufficiently enhanced by a moderate *-I* effect (as in the case of Cl), allowing the radical to persist long enough to participate in SWCNT functionalization. The other substituents, *i.e.*, 4-COOCH$_3$ ($\sigma = 0.450$), 4-Cl ($\sigma = 0.227$), and 3-CN ($\sigma = 0.560$), exhibit significantly lower reactivity, with 4-COOCH$_3$ being slightly more reactive than the other two. The relatively higher reactivity of 4-COOCH$_3$ is most likely due to its strong resonance effect, which plays a less significant role in the reactivity of radical initiators in the 3- position. Regarding the values of full width at half-maximum (FWHM) of the $E_{11}$* defect peaks, they remain consistent across all initiators after normalization (Figure 4d), indicating a similar type of defect formation irrespective of the initiator structure. Importantly, the PL spectra of SWCNTs functionalized using BPO derivatives in many cases did not change even after months of storage, underscoring their robust nature. Such good stabilization was achieved by using a low initiator concentration and a high reaction temperature. A detailed description is provided in SI, section 2.6.

To better understand the process of defect implementation, we also conducted Raman spectroscopy for selected modified (6,5) SWCNTs. The intensity ratios of D to G bands were calculated to assess the extent of SWCNT modification (Table S3). Then, the area ratios of $E_{11}$*/$E_{11}$ obtained for these samples were plotted against the $I_D/I_G$ to compare the effectiveness of various substituted BPOs in imprinting the photoluminescence defects. For SWCNTs functionalized at 70 °C, low $E_{11}$*/$E_{11}$ area ratios were noticed, but the $I_D/I_G$ ratios indicative of the relative content of defects were high (Figure 4e), suggesting an excessive amount of non-emissive defect states. Even 4-CH$_3$ substituted BPO, which did not exhibit the $E_{11}$* defect feature, contained a considerable number of defects ($I_D/I_G$ ≈ 0.32 for [R-BPO]/[CNT] = 84 as well as 168). SWCNTs functionalized in a higher temperature range showed elevated values of $E_{11}$*/$E_{11}$ area ratio in relation to $I_D/I_G$ ratios. Interestingly, SWCNTs functionalized at 100°C with 4-OCH$_3$-BPO displayed both low $E_{11}$*/$E_{11}$ and $I_D/I_G$ ratios even for overwhelming excess of the initiator [R-BPO]/[CNT] = 335, which indicates that either the radicals created in this reaction did not react effectively with the SWCNTs or the initiator remained largely intact. This latter possibility is improbable since the electron-donating effect of the methoxy group in 4-OCH$_3$-BPO is expected to cause significant dipole-dipole repulsion within the peroxy group, which dramatically decreases the thermal stability of the radical initiator [28]. Moreover, we discovered that with increasing the Hammett constant of the substituent, the same $I_D/I_G$ corresponded to higher $E_{11}$*/$E_{11}$ ratios (Table S3). Based on this finding, we postulate that such reactants offer more effective implantation of the photoluminescent defects into SWCNTs.

Furthermore, while the influence of the substituent present in the moiety attached to the SWCNT on the optical properties of SWCNTs has been reported [12,15,45], it is usually discussed only in the context of the optical trap depth it creates. At the same time, the possible impact of this factor on the propensity for SWCNT modification remains unknown. Considering that the electron-poor substituents better improve the PLQY of SWCNTs [10], identifying the origins of this phenomenon is essential. To elucidate this, we subjected SWCNTs to chemical modification with 3-Cl-BPO, 3-F-BPO, and 4-F-BPO at 100°C for 1 h and then analyzed the samples by Raman spectroscopy. Different positional isomers and two types of halogens were applied to tune the reactivity of the BPO derivatives. 4-F-BPO turned out to be the least reactive compound, giving an $I_D/I_G$ ratio of 0.06 (Figure 5a). As previously discussed, substituents in the 4- position are unable to deplete the electron density of the aromatic parts of BPO to a sufficient extent to promote the homolytic scission of this molecule, which is necessary to generate radicals for SWCNT modification. Consequently, the $I_D/I_G$ ratio of 0.06 indicated a negligible degree of chemical modification of SWCNTs with 4-F-BPO, which was not much greater than that of pristine SWCNTs (0.02).

Hence, 3-isomers of halogenated BPO, the electronic configuration of which is more appropriate, were employed. As expected, 3-F-BPO functionalized performed better, and an $I_D/I_G$ ratio of 0.18 was recorded. Furthermore, the chlorine-containing analog of this compound, i.e., 3-Cl-BPO, made a much more substantial impact on the surface of SWCNTs, reaching an $I_D/I_G$ ratio of 0.41. At the same time, the corresponding PL spectrum was of poor quality (Figure 5b), confirming a considerable modification of the material. In light of the foregoing, the obtained results strongly suggest that treatment of SWCNTs with 3-Cl-BPO not only introduces deeper traps for mobile excitons (Figure 5cd), compared to the exposure to 4-F-BPO but also impacts the underlying photophysics. We hypothesize that the reported in the literature improvement of PLQY and $E_{11}*/E_{11}$ ratios related to the attachment of electron-withdrawing substituents to the SWCNT side wall [10] is likely caused by the increased amount of such groups on the SWCNT surface. The density of luminescent defects in SWCNTs is directly related to the $I_D/I_G$ ratios measured by Raman spectroscopy [5,46] and the highest $E_{11}*/E_{11}$ ratios were recorded for the most severely affected SWCNTs, which were exposed to the electron-poor reactants containing substituents in the 3-position. Thus, in addition to the previously disclosed increased exciton trap depth for electron-deficient SWCNT addenda [47], it should also be considered what happens when an exciton experiences thermal detrapping [48]. For these SWCNTs, which are supposedly more functionalized, it is likely that excitons, which generally have high mobility [49,50], can become effectively retrapped by the defect sites present in abundance. Consequently, radiative recombination through the defect channel becomes much more probable, which explains why the $E_{11}*$ feature is much more pronounced in as-modified SWCNTs and why these very SWCNTs display leading PLQYs.

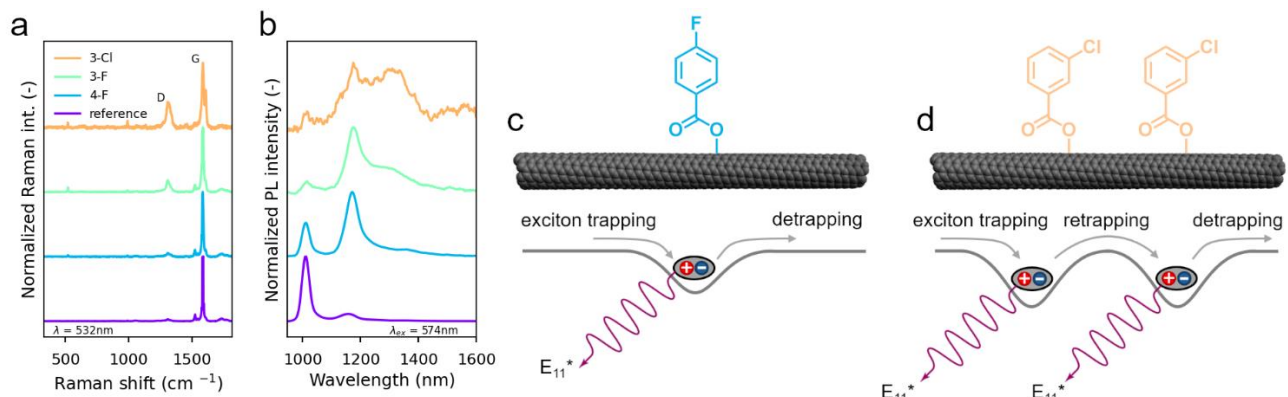

**Figure 5** a) Raman spectra of pristine SWCNTs and modified with 3-Cl-BPO, 3-F-BPO, and 4-F-BPO, b) corresponding PL spectra normalized to the intensity of the maximum of the peak of the highest intensity, SWCNT reacted with c) 4-F-BPO, and d) 3-Cl-BPO with the schematics of the main phenomena related to exciton dynamics in SWCNTs modified with luminescent defects.

### 3.6. Relationship between the $E_{11}$* peak position and the Hammett substitution constants

Upon closer examination of the $E_{11}$* peak positions for various substituents, as exemplified in Figure 4f, subtle variations in peak location are discernible depending on the substituent type and its position (3- vs. 4-). Considering that the functionalization approach involving the attachment of a functional group to SWCNTs via a carbonyl linker (SWCNT-O-C(O)-Ph-R) has not been previously investigated in this context, we aimed to address this gap and relate our findings to existing literature data for SWCNT-Ph-R type defects. For the analysis, we selected the spectra obtained with low concentrations of the initiator and exhibiting similar defect densities (Figure S14). The $E_{11}$*/$E_{11}$ area ratios, as well as the positions of the peaks, were calculated using a model visualized in Figure S4. Values obtained for all used peroxides were collected in Table S4. $E_{11}$* peaks' intensities were normalized to show their exact positions and reveal that increasing the Hammett substituent constant indeed caused further redshift of the defect peak in PL spectra of modified SWCNTs (Figure 4f). The relation between the Hammett substituent constant and the spectral shift (difference in spectral positions between $E_{11}$* and $E_{11}$ in nm, often referred to as the optical trap depth) is shown in Figure 4g.

Our investigation into the effect of substituted benzoyloxy radicals on the photoluminescence of (6,5) SWCNTs reveals a complex interplay between substituent electronic properties, reaction temperature, and the resulting optical spectra modifications. In agreement with the general trend reported by Piao et al. [13] for aryl-functionalized SWCNTs dispersed by surfactants in water, we also observe progressing redshift of the $E_{11}$* peak with increasing σ, linked with electron donating or withdrawing effect of substituents (Figure 4f). This observation is consistent with the expected lowering of the LUMO energy level upon the introduction of electron-withdrawing functional groups. However, in contrast to the perfectly linear correlation reported by Piao et al., our data demonstrate differences in this aspect, forming small groups with different Hammett constants but the similar $E_{11}$* redshift, e.g. 4-CH$_3$, 3-CH$_3$, -H, 4-F and 3-OCH$_3$ (Figure S15) or 4-Cl, 3-F, 3-COOCH$_3$ or 3-NO$_2$, 4-NO$_2$, 4-CN. This non-linearity is particularly evident at both low and high values of the Hammett substituent constants.

Furthermore, we observed noteworthy differences in the spectral shifts induced by 3- and 4-isomers (Table S4). For instance, the $E_{11}$*-$E_{11}$ shift for 3-F and 4-F at 100°C differs by 5 nm (160 nm vs 155 nm) and for the ester group (160 nm vs 165 nm). A minor $E_{11}$*-$E_{11}$ shift difference (2 nm) was observed for SWCNTs modified with chlorine and cyano-substituted BPO, while methyl derivatives resulted in the same shift. This isomer-dependent behavior suggests that the position of the substituent on the benzoyloxyl radical subtly influences the local defect environment, which impacts the SWCNT electronic structure. These differences could arise from steric effects, variations in the spatial arrangement or orientation of the attached functional group relative to the SWCNT surface (potentially influenced by interactions within the conjugated polymer wrapping), or subtle changes in the electronic coupling between the substituent and the SWCNT lattice, depending on the meta or para orientation.

It is important to note that this observation is not entirely without precedent. Indeed, our previous studies on solvatochromism in polymer-dispersed SWCNTs have demonstrated the influence of solvents with varying dielectric properties. In those cases, solvent polarity, manifested as changes in the dielectric constant, was shown to interact with the polymer wrapping and induce additional strain effects. Consequently, the observed $E_{11}$ peak shift was attributed to the combined influence of dielectric and strain interactions [51]. This indicates that the $E_{11}$* defect peak shift observed in our study, induced by the attachment of moieties with differing dielectric properties, may also be mediated by alterations in the polymer wrapping of the SWCNT surface, thereby modulating its contribution to the observed spectral shift. However, polymer-free SWCNTs functionalized in water, as reported by Shiraki et al. [15], also exhibit a different $E_{11}$*-$E_{11}$ shift depending on the type of positional isomer used for their covalent modification, supporting that the observed effect is the result of the direct influence of the attached moiety on the SWCNT.

In summary, our findings demonstrate that while the electronic properties of benzoyloxyl radicals, as characterized by Hammett substituent constants, play a crucial role in tuning the $E_{11}$* redshift, the relationship is not entirely linear and is significantly modulated by substituent position and to a lesser extent by the reaction temperature. This behavior highlights the complex interplay of factors governing SWCNT functionalization in polymer-organic solvent systems and underscores the potential for fine-tuning SWCNT optical properties by careful control of radical chemistry and reaction conditions.

### 3.8. Low-temperature (6,5) SWCNT functionalization with BPO derivatives

Given the sufficient radical activity at room temperature after thermal initiation, we aimed to examine if a further significant temperature decrease from 70°C to 40°C could extend the process duration and provide enhanced control. Functionalization at low temperatures presents a considerable challenge due to slower decomposition kinetics and potentially reduced radical reactivity. To assess the feasibility of such modifications, we investigated the functionalization of (6,5) SWCNT at 40°C over an extended reaction period (24 h), using a high radical-to-SWCNT molar ratio of 251:1. The results, shown in Figure 6a-b, provide insights into how different benzoyloxyl radicals and their derivatives influence defect formation under these conditions.

Among the tested initiators, the electron-rich 4-OCH$_3$-BPO remained inactive, failing to introduce detectable defects. In contrast, both chloro-substituted derivatives (3-Cl and 4-Cl) successfully generated $E_{11}$* defects, albeit with distinct spectral features. Notably, 3-Cl-BPO produced broader defect-related peaks, resembling previous high-temperature experiments where excessive radical concentrations led to non-selective modifications, formation of defects with various morphologies, or defects accumulation in close vicinity to already installed moieties. A possible solution to this issue could be functionalization with a lower molar excess of these initiators while extending the reaction time to several days.

For 4-CN-BPO and unsubstituted BPO, the defect-induced peaks were among the narrowest recorded, indicating a more uniform functionalization pattern when the treatment was executed at a low temperature. However, this selectivity came at the expense of defect density, as indicated by the $E_{11}^*/E_{11}$ area ratio, which reached ca. 2 for 4-CN-BPO and only ca. 1 for BPO. These findings suggest that while low-temperature functionalization enables the selective formation of defects, it necessitates prolonged reaction times and high radical concentrations, resulting in a trade-off between defect selectivity and density.

### 3.9. (7,5) SWCNT functionalization

The inherent differences in reactivity among BPO derivatives offer a powerful tool for achieving covalent functionalization of SWCNTs with various chiralities and, thus, reactivities. The significance of this capability may be used to enable the generation of $E_{11}^*$ emission that is further redshifted towards the telecommunication band compared to functionalized (6,5) SWCNTs. While functionalization generally induces PL redshift and the emergence of new $E_{11}^*$ peaks, the magnitude and nature of this spectral shift are dependent on the specific SWCNT chirality being modified [12]. Furthermore, SWCNT reactivity is known to substantially decrease with increasing SWCNT diameter because of reduced curvature and bandgap variation [52,53].

As previously demonstrated, the functionalization process often does not terminate spontaneously upon initiation. Therefore, precise control over initiator reactivity, its tunability, and general performance becomes paramount. A crucial goal of achieving such control is to avoid detrimental over-functionalization, a condition known to quench the fluorescence of SWCNTs. Literature sources indicate that an optimal defect density range for enhanced photoluminescence is approximately 5-10 defects per 1000 nm [47]. Leveraging the tunable reactivity profiles of BPO derivatives presents a compelling strategy for achieving chirality-specific functionalization. For instance, the high reactivity of 3-NO$_2$-BPO can be effectively harnessed for the functionalization of (7,5) chirality, which is more challenging to modify due to increased diameter (d=0.829 nm) and consequently lower reactivity compared to (6,5) SWCNTs (d=0.757 nm). Figure 6c,d provides a clear illustration of this principle. Additionally, other than 3-NO2-BPO, less reactive BPO derivatives, such as 3-Cl-BPO and 4-Cl-BPO, provide a route to controlled functionalization of (7,5) SWCNTs, allowing for fine adjustments at low defect densities. Figure 6c displays this approach, showcasing functionalization using BPO, 4-COOCH$_3$-BPO, 3-Cl-BPO, 4-Cl-BPO, 3-CN-BPO, and 3-NO$_2$-BPO at varied temperatures (70°C and 100°C) and concentrations (Figure S16). Spectra obtained after 4-5 hours of reaction (Figure 6d) highlight the concentration-dependent functionalization using $3\text{-}NO_2\text{-}BPO$, with normalized spectra revealing a progressive increase in $E_{11}*/E_{11}$ peak area ratio and peak broadening at 1300-1320 nm. This controlled methodology, utilizing BPO derivatives with tailored reactivity, offers a versatile toolkit for precise engineering the optical properties of even the less reactive SWCNTs for diverse applications.

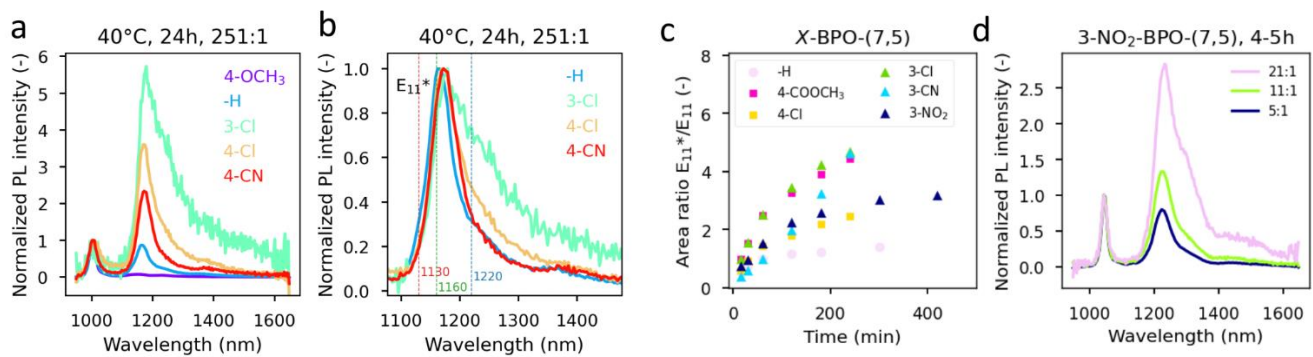

**Figure 6** PL spectra of (6,5) SWCNTs functionalized using BPO radicals substituted with groups listed in the legends, normalized to a) $E_{11}$ and b) $E_{11}*$ PL intensity. Functionalization was conducted for a long time (24 h) at a low temperature (40°C). A large molar ratio of initiator to SWCNTs (251:1) was necessary for the $E_{11}*$ to emerge in these conditions. Spectra were obtained for 574 nm excitation wavelength. c) $E_{11}*/E_{11}$ area ratios measured for (7,5) SWCNTs reacted with differently substituted BPO derivatives at 100°C (except for 3-CN-BPO, which was reacted at 70°C). For better visibility, values obtained using 3- and 4-substituted BPOs were plotted using triangles and squares, respectively. d) Exemplary spectra of the (7,5) SWCNTs modified using different [R-BPO]/[CNT], obtained after 4-5 hours of reaction at 100 °C, normalized to $E_{11}$ peak intensity. Spectra were obtained for 653 nm excitation wavelength.

## 4. Conclusions

Conducted investigations present a comprehensive view into the covalent functionalization of (6,5) SWCNTs (and in limited scope also (7,5) chirality) using a library of BPO derivatives with various substituents. Our findings highlight the unique control that can be achieved over the optical properties of SWCNTs through meticulous manipulation of radical chemistry and reaction parameters. A systematic exploration using a variety of self-synthesized BPO derivatives, including changes in the electron character of the substituent, steric hindrance, and positional isomerism (3- vs. 4-), elucidated the complex interaction between initiator structure and SWCNT functionalization efficiency.

In particular, the study showed that the electronic and steric properties of the substituted benzoyloxyl radical have a profound effect on SWCNT functionalization. The reactivity of various BPO derivatives was found to be highly dependent on the type of substituent, with electron-withdrawing groups generally increasing reactivity in both 3- and 4- positions. The 4-position of the substituent introduces more nuanced effects related to resonance stabilization. The carried-out investigations allowed for the identification of both highly reactive initiators, such as those containing $3\text{-}NO_2$ or 3-Cl substituents, and less reactive counterparts, such as those containing $3\text{-}CH_3$ or 4-F groups, enabling a tunable range of defect densities and spectral modifications. By exploiting the difference in reactivity of BPO derivatives, chirality-specific functionalization becomes a tangible goal, opening up the possibility of tailoring the optical properties of SWCNTs and potentially extending them toward the telecommunication window. Furthermore, the successful implementation of this methodology to the functionalization of (7,5) SWCNTs, which, due to its lower curvature, experiences lower reactivity, demonstrates the versatility of the developed synthetic toolbox. Interestingly, this study reveals that the SWCNTs modified with certain electron-poor substituents are more likely to experience exciton retrapping, which rationalizes why, in the literature, such-modified SWCNTs offer superior PLQY values and $E_{11}^*/E_{11}$ ratios.

Additionally, our study highlights the key role of reaction conditions, especially temperature and initiator concentration, in achieving controlled functionalization. Kinetic studies have demonstrated that defect density can be precisely modulated, with lower initiator concentrations resulting in the complete depletion of radicals during the process and stable optical spectra for extended periods. Exploring functionalization at low temperatures (40°C) further enhances the control regime, albeit at the expense of a trade-off between defect density and reaction time. Moreover, we explored strategies to terminate the functionalization process, including the use of radical scavengers. While TEMPO showed some promise in limiting further defect formation, precise control of initiator concentration proved to be a more reliable and cleaner approach to achieving stable functionalized SWCNT dispersion. The long-term stability of optical properties of such dispersions, even without post-reaction purification steps, further emphasizes the practical feasibility of this methodology.

The observed correlation between the $E_{11}^*$ peak redshift and the Hammett substitution constants, although not strictly linear, provides a valuable framework for predicting and rationalizing the optical properties of functionalized SWCNTs. Differences in the value of the shift, compared to already published studies, highlight the system's complexity and provide insight into the interplay of electronic effects, such as steric hindrance, spatial arrangement at the defect site, and polymer wrapping.

In summary, our research provides the first comprehensive understanding of SWCNT functionalization based on modification of benzoyloxy radical reactivity, offering a novel approach for the controlled engineering of luminescent defects. The presented systematic investigation, detailed kinetic studies, and exploration of substituent effects provide valuable insights into the surface chemistry of SWCNTs, paving the way for the rational design of nanomaterials for various optoelectronic applications. In parallel, the library of BPO derivatives synthesized in-house uncovered how, through modification of their structure, one can control the degree of their electrophilicity and, thus, the effectiveness of many fundamental chemical transformations in which BPO is regarded as an indispensable component.

**Conflicts of interest**

There are no conflicts of interest to declare.

**Acknowledgments**

The authors would like to thank the National Science Centre, Poland (under the SONATA program, Grant agreement UMO-2020/39/D/ST5/00285) for supporting the research and Metropolis GZM, Poland (under Metropolitan Science Support Fund, Grant agreement No. RW/61/2025) for supporting the open access publication of the results of this study. The authors would also like to thank Marta and Alicja Kalyta for their help with preparing multiple functionalization reactions and Dominik Just for his help with Raman spectra measurements.

(Supporting Information)

# Programing optical properties of single-walled carbon nanotubes with benzoyl peroxide derivatives of tailored chemical characteristics

Andrzej Dzienia [a, *], Patrycja Taborowska [a], Paweł Kubica-Cypek [a], Dawid Janas [a,*]

[a] Department of Chemistry, Silesian University of Technology, B. Krzywoustego 4, 44-100, Gliwice, Poland

* Corresponding authors: Andrzej.Dzienia@polsl.pl, Dawid.Janas@polsl.pl

**Contents**

## 1. Experimental

### 1.1. Materials

All chemical reagents, i.e., 9,9-dioctyl-2,7-dibromofluorene (Angene, cat. number: AG002BU7, CAS: 198964-46-4, purity: 98%), 9,9-dioctylfluorene-2,7-bis(boronic acid pinacol ester) (Angene, cat. number: AG0034EZ, CAS: 196207-58-6, purity: 98%), Aliquat 336 TG (Alfa Aesar, cat. number: A17247, CAS: 63393-96-4, purity N/A), tetrakis (triphenylphosphine)palladium - Pd(PPh3)4 (Apollo Scientific, cat. number: OR4225, CAS: 14221-01-3, purity: >99%), 3-nitrobenzoyl chloride (Acros, cat. number: 155840050, CAS: 121-90-4, purity: 98%), 4-nitrobenzoyl chloride (Alfa Aesar, cat. number: A12543.09, CAS: 122-04-03, purity: 98%), 3-chlorobenzoyl chloride (Apollo Scientific, cat. number: OR4528-25g, CAS: 618-46-2, purity: 99%), 4-chlorobenzoyl chloride (Alfa Aesar, cat. number: A16325.14, CAS: 122-01-0, purity: 98% 4-cyanobenzoyl chloride (Apollo Scientific, cat. number: OR3784-5g, CAS: 6068-72-0, purity: tech), 4-anisoyl chloride (Acros, cat. number: 104850050, CAS: 100-07-2, purity: 99%), 3-anisoyl chloride (Acros, cat. number: 212970050, CAS: 1711-05-3, purity: 99%), 3-methylbenzoyl chloride (AmBeed, cat. number: A895978, CAS: 618-48-4, purity: 97%) 4-toluoyl chloride (AmBeed, cat. number: A157814, CAS: 874-60-2, purity: 98%), 3-(Methoxycarbonyl)benzoic acid (AmBeed, cat. number A269264, CAS: 1877-71-0, purity: 97%), 4-(Methoxycarbonyl)benzoic acid (AmBeed, cat. number A261585, CAS: 1679-64-7, purity: 98%), 4-Fluorobenzoic acid (Angene, cat. number AG00341D, CAS 456-22-4, purity: 98%) 3-Fluorobenzoic acid (AmBeed, cat. number A160641, CAS 455-38-9, purity: 97%), 3-Cyanobenzoic acid (AmBeed, cat. number A228622, CAS 1877-72-1, purity: 98%), 4-nitrobenzoic acid (Thermo Scientific, cat. number A12543.09, CAS 62-23-7, purity: 98%), oxalyl chloride (Alfa Aesar, cat. number: A18012.18, CAS: 79-37-8, purity: 98%), thionyl chloride (Acros, cat. number: A0401657, CAS: 7719-09-7, purity: 99.7%), sodium hydroxide (PPH Stanlab, cat. number: 011-002-00-6, CAS: 1310-73-2, purity: pure p. a.), hydrogen peroxide (PPH Stanlab, cat. number: 008 003-029, CAS: 7722-84-1, purity: 30%), magnesium sulfate anhydrous (Chempur, cat. number: 116137606, CAS: 7487-88-9, purity: pure p. a.), diethyl ether (Thermo Scientific, cat. number: 364335000, CAS: 60-29-7, purity: 99,5%, extra dry over molecular sieves, stabilized with BHT AcroSeal®), dichloromethane (Acros, cat. number: 348460010, CAS: 75-09-2, purity: 99.8% Extra Dry over Molecular Sieve, Stabilized, AcroSeal®), N,N-dimethylformamide (Acros, cat. number: 35843100, CAS: 68-12-2, purity: 99.8%, extra dry over molecular sieves AcroSeal®) and toluene (Alfa Aesar, cat. number: 19376.K2, CAS: 108-88-3, spectrophotometric grade, purity: >99.7%), were used as supplied, without additional purification or drying. The study was carried out using (6,5)-enriched CoMoCAT SG65i SWCNTs (Sigma-Aldrich, product number: 773735, lot: MKCM5514, purity: 95%-carbon basis).

### 1.2. Characterization

1.2.1.    Nuclear Magnetic Resonance ($^1$H NMR)

The proton Nuclear Magnetic Resonance ($^1$H NMR) spectra were typically registered using an Agilent 400-MR spectrometer operating at 400 MHz with CDCl$_3$ or toluene-d$_8$ (as the solvent. DMSO-d$_6$ was used for more polar BPO derivatives, i.e. 4-OCH$_3$, 4-NO$_2$ and 4-CN. The $^1$H chemical shifts were recorded in δ (ppm), utilizing the residual peak of the deuterated solvents as a reference. Standard experimental conditions were applied (64 scans, T=20°C). The spectrometer was equipped with a OneNMR probe, ProTune automatic probe tuning, and VnmrJ 3 software. Subsequent spectral processing was conducted using the Mestrenova software suite.

1.2.2.    Size Exclusion Chromatography (SEC)

Molecular weights and dispersity (Đ) indices were determined through SEC employing an Agilent 1260 Infinity system (Agilent Technologies), which was equipped with an isocratic pump, autosampler, degasser, thermostatic box for columns, and a differential refractometer, MDS RI Detector. Data collection and processing were performed using Agilent Technologies' Addon Rev. B.01.02 data analysis software. SEC-calculated molecular weights were derived from calibration using linear polystyrene standards ranging from 580 to 300,000 g mol$^{-1}$. The separation process involved a pre-column guard (5 µm, 50 × 7.5 mm) and two columns, namely PLGel 5 µm MIXED-C (300 × 7.5 mm) and PLGel 5 µm MIXED-D (300 × 7.5 mm). Measurements were conducted using chlorobenzene (HPLC grade) as the solvent at a temperature of 30 °C and a flow rate of 0.8 mL min$^{-1}$.

### 1.3. Synthesis of PFO-BPy6,6' and PFO to purify SWCNTs

The PFO-BPy6,6' and PFO polymer was synthesized in-house according to the earlier described procedure [1]. In brief, the Suzuki polycondensation was performed using 9,9-dioctylfluorene-2,7-diboronic acid bis(pinacol)ester and 6,6'-dibromo-2,2'-bipyridine (for PFO-BPy6,6') or 9,9-dioctyl-2,7-dibromofluorene (for PFO) monomers in toluene and 1M $Na_2CO_3$ aqueous solution, with Aliquat 336 as a phase transfer catalyst and $Pd(PPh_3)_4$ as a catalyst, under argon atmosphere at 85°C for 5 days.

### 1.4. Synthesis of BPO derivatives as radical substrates for covalent SWCNT functionalization

1.4.1.    General synthesis of acid chlorides with thionyl chloride

Initially, 2 g of the appropriate benzoic acid derivative was dissolved in 10 mL of toluene (approximately 5-10 mL per 1 g of benzoic acid derivative) in a round-bottom flask. Anhydrous N,N-dimethylformamide (DMF, 2 drops) was added as a reaction catalyst. Subsequently, thionyl chloride was added dropwise to the flask at a molar ratio of 12:1 (thionyl chloride to benzoic acid derivative) under continuous stirring. The reaction mixture was then heated to 80°C for approximately 2 h. Excess thionyl chloride was removed via distillation using a scrubber containing a concentrated NaOH solution. The resulting residue was washed twice with diethyl ether (2 x 5 mL) and once with toluene (1 x 5 mL). Finally, the solvents were removed by rotary evaporation under reduced pressure. Employing this method based on thionyl chloride, the following substituted benzoic acid chlorides were synthesized: 4-chloro and 4-nitrobenzoic acid chloride.

1.4.2.    General synthesis of acid chlorides with oxalyl chloride

First, 2 g of the appropriate benzoic acid derivative was dissolved in 25 mL of anhydrous dichloromethane. Anhydrous N,N-dimethylformamide (DMF, 2 drops) was added as a reaction catalyst. Subsequently, oxalyl chloride was added dropwise to the flask at a molar ratio of 1.5:1 (oxalyl chloride to benzoic acid derivative) under continuous stirring. After bubbling ceased, an additional 0.25 molar equivalents of oxalyl chloride were added to the mixture, and stirring was continued for a further 15 min to ensure completion of the reaction. Excess oxalyl chloride and dichloromethane (DCM) were then removed by rotary evaporation at 40 °C under reduced pressure. Utilizing this method based on oxalyl chloride, the following substituted benzoic acid chlorides were synthesized: 4-fluoro-, 3-fluoro-, 3-cyano-, 4-(methoxycarbonyl)-, and 3-(methoxycarbonyl)benzoic acid chlorides. For other derivatives not listed, specifically 3-chloro-, 4-cyano-, 3-nitro-, 4-methoxy-, 3-methoxy-, 4-methyl-, and 3-methylbenzoic acid chlorides, commercially available acid chlorides were used.

### 1.4.3. General synthesis of BPO derivatives

Approximately 2 g of the appropriate acid chlorides were diluted in 1 mL of diethyl ether or DCM in a round bottom flask and cooled with cold water bath. Then 30 wt.% aqueous hydrogen peroxide solution (0.57 molar equivalents) was added dropwise to the mixture over 10 minutes. This was followed by the dropwise addition of the 1.25 molar equivalents of aqueous NaOH solution over 20 minutes. Then the biphasic reaction mixture was left stirring for 1 h. The only exceptions to general synthesis were for alkyl substituents. These reactions had to be left to stir for 24 hours. If the product was a precipitate, it was collected by filtration, then washed with distilled water (2 x 5 mL) and diethyl ether / DCM (1 x 5 mL). However, if the product was present in the organic phase, the layers were separated using separatory funnel, washed with water (3 x 5 mL), and dried with anhydrous $MgSO_4$. Then the solvent was evaporated using a rotatory evaporator at 40 °C under reduced pressure.

## 2. Results and discussion

### 2.1. Characterization of the synthesized compounds

The molecular weight characteristics of the PFO-BPy6,6' and PFO were determined using SEC. This analysis revealed weight-average molecular weights ($M_w$) of 98.5 kg mol$^{-1}$ and 34.0 kg mol$^{-1}$, number-average molecular weights ($M_n$) of 35.0 kg mol$^{-1}$ and 10.2 kg mol$^{-1}$, and polydispersity indices (Đ) of 2.82 and 3.32 kg mol$^{-1}$ for PFO-BPy6,6' and PFO, respectively.

The [1]H NMR spectra of in-house synthesized PFO-BPy6,6' and PFO are presented in Figure S1a and S1b. The observed chemical shifts are consistent with literature values [2] and analogous to those obtained in our previous studies [3–6]. The [1]H NMR spectra of BPO (Figure S2) and its derivatives were recorded and are analyzed in the subsequent sections. For comparative purposes, the experimental chemical shift values for BPO derivatives are compiled and presented in Table S1. For most of the synthesized BPO derivatives, literature values for chemical shifts were identified and are listed in a separate column of Table S1, along with citations to the corresponding publications. Where literature references were available, the experimentally obtained NMR data exhibited good agreement with the reported values.

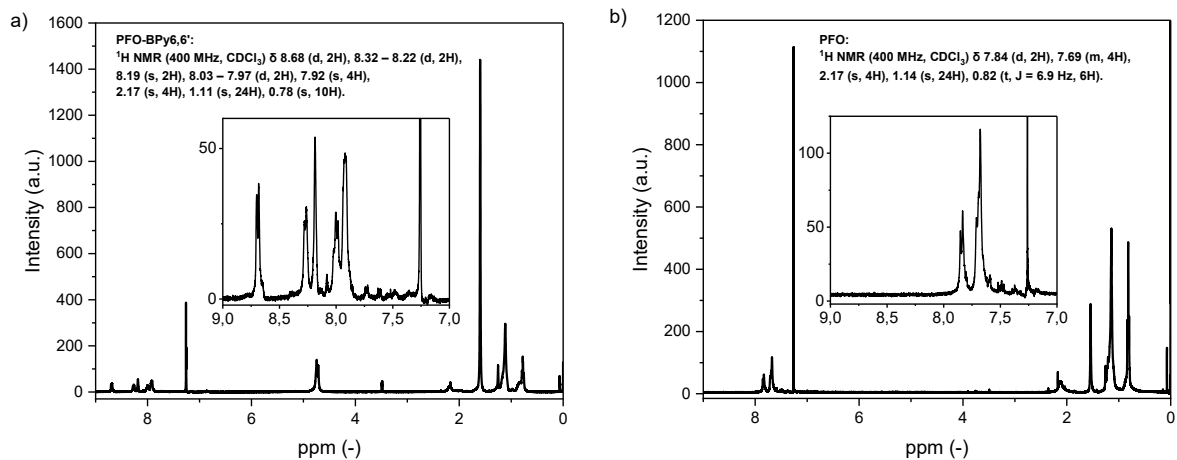

**Figure S1** $^1$H NMR spectra of a) PFO-BPy6,6' and b) PFO synthesized in-house.

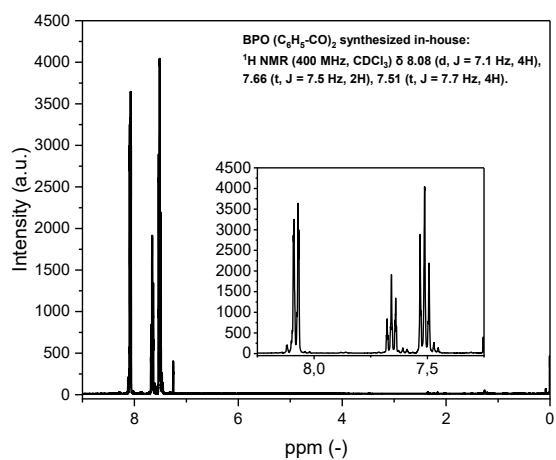

**Figure S2** $^1$H NMR spectrum of benzoyl peroxide (BPO) synthesized in-house.

**Table S1** Experimental and literature $^1H$ NMR chemical shifts for BPO and its derivatives

| Name | $^1H$ NMR shifts (measured) | $^1H$ NMR shifts (literature) | Ref. |
|---|---|---|---|
| -H | 8.08 (d, $J$ = 7.1 Hz, 4H), 7.66 (t, $J$ = 7.5 Hz, 2H), 7.51 (t, $J$ = 7.7 Hz, 4H) | 8.08 (d, $J$ = 8.3 Hz, 4H), 7.71 −7.62 (m, 2H), 7.52 (t, $J$ = 7.7 Hz, 4H) | 7 |
| 4-CH₃ | 7.96 (d, $J$ = 8.2 Hz, 4H), 7.30 (d, $J$ = 8.0 Hz, 4H), 2.44 (s, 6H) | 7.97 (d, $J$ = 8.0 Hz, 4H), 7.31 (d, $J$ = 8.0 Hz, 4H), 2.44 (s, 6H) | 7 |
| 3-CH₃ | 7.90 − 7.85 (m, 4H), 7.46 (d, $J$ = 7.3 Hz, 2H), 7.40 (dd, $J$ = 11.4, 4.3 Hz, 2H), 2.43 (s, 6H) | 7.94 − 7.81 (m, 2H), 7.52 (d, $J$ = 7.6 Hz, 1H), 7.43 (t, $J$ = 7.7 Hz, 1H), 2.44 (s, 3H) * | 8 |
| 4-OCH₃ | DMSO-d6: 7.92 − 7.89 (m, 4H), 7.01 (d, $J$ = 8.9 Hz, 4H), 3.82 (s, 6H) | 8.03 (d, J= 9.0 Hz, 4H), 6.98 (d, $J$ = 9.0 Hz, 4H), 3.89 (s, 6H) | 7 |
| 3-OCH₃ | 7.69 − 7.65 (m, 2H), 7.57 (dd, $J$ = 2.6, 1.5 Hz, 2H), 7.42 (t, 2H), 7.20 (ddd, $J$ = 8.4, 2.7, 1.0 Hz, 2H), 3.87 (s, 6H) | 7.67 (d, J= 8.0 Hz, 2H), 7.60 −7.54 (m, 2H), 7.42 (t, J= 8.0 Hz, 2H), 7.20 (dd, J= 8.0, 2.4 Hz, 2H), 3.87 (s, 6H) | 7 |
| 4-COOCH₃ | 8.17 (q, $J$ = 8.6 Hz, 8H), 3.98 (s, 6H) | Not found | - |
| 3-COOCH₃ | 8.74 (t, $J$ = 1.5 Hz, 2H), 8.39 − 8.32 (m, 2H), 8.28 − 8.25 (m, 2H), 7.64 (t, $J$ = 7.8 Hz, 2H), 3.98 (s, 6H) | Not found | - |
| 4-F | 8.10 (dd, $J$ = 9.0, 5.3 Hz, 4H), 7.20 (dd, $J$ = 8.9, 8.4 Hz, 4H) | 8.10 (dd, $J$ = 8.6, 5.3 Hz, 4H), 7.20 (t, $J$ = 8.6 Hz, 4H) | 7 |
| 3-F | 7.88 (d, $J$ = 7.8 Hz, 2H), 7.79 − 7.74 (m, 2H), 7.56 − 7.48 (m, 2H), 7.42 − 7.35 (m, 2H) | Not found | - |
| 4-Cl | 8.01 (d, $J$ = 8.8 Hz, 2H), 7.50 (d, $J$ = 8.8 Hz, 2H) | 8.01 (d, $J$ = 8.5 Hz, 4H), 7.50 (d, $J$ = 8.5 Hz, 4H) | 7,9 |
| 3-Cl | 8.05 (t, $J$ = 1.7 Hz, 2H), 7.98 − 7.94 (m, 2H), 7.65 (ddd, $J$ = 8.1, 2.1, 1.0 Hz, 2H), 7.47 (t, $J$ = 8.0 Hz, 2H). | 7.96 (d, $J$ = 7.8 Hz, 2H), 7.65 (d, $J$ = 8.0 Hz, 2H), 7.47 (t, $J$ = 7.8 Hz, 2H) * | 7,9 |
| 4-CN | CDCl₃ / toluene-d8: 8.19 (d, $J$ = 8.2 Hz, 4H), 7.76 (d, $J$ = 8.2 Hz, 4H) DMSO-d6: 8.07 (d, $J$ = 8.2 Hz, 1H), 7.97 (d, $J$ = 8.5 Hz, 1H) | 8.16 (d, $J$ = 8.5 Hz, 2H), 7.84 (d, $J$ = 8.6 Hz, 2H) * | 9 |
| 3-CN | 8.40 (dd, $J$ = 9.7, 8.7 Hz, 2H), 8.31 (dd, $J$ = 8.0, 1.2 Hz, 2H), 7.98 (d, $J$ = 8.0 Hz, 2H), 7.71 (t, $J$ = 7.9 Hz, 2H) | Not found | - |
| 4-NO₂ | DMSO-d6: 8.30 (d, $J$ = 8.2 Hz, 4H), 8.20 (d, $J$ = 8.2 Hz, 4H) | 8.37 (d, $J$ = 8.8 Hz, 2H), 8.25 (d, $J$ = 8.7 Hz, 2H) * | 9,10 |
| 3-NO₂ | 8.93 (dd, $J$ = 2.2, 1.7 Hz, 2H), 8.56 (dd, $J$ = 8.3, 2.3 Hz, 2H), 8.42 (d, $J$ = 7.8 Hz, 2H), 7.80 (t, $J$ = 8.0 Hz, 2H) | Not found | - |

Peroxides were measured in CDCl₃ unless stated otherwise. * The total hydrogen count is determined by multiplying by 2, as the compound's symmetry allows for analysis of either one side or the entire molecule.

## 2.2. Characterization of the isolated SWCNTs used for functionalization

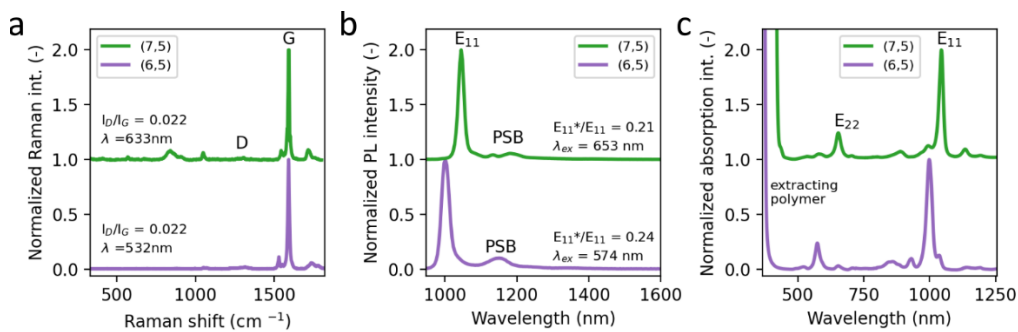

**Figure S3** Spectral characteristics of the starting SWCNT material before functionalization: a) Normalized Raman spectra of drop-casted pristine SWCNTs, with G and D bands. b) PL spectra with original $E_{11}$ peaks and photoluminescence sidebands (PSBs) [11], c) $E_{11}$-normalized absorption spectra with $E_{11}$ and $E_{22}$ excitonic bands and bands originating from the extracting polymers (PFO-BPy6,6' in case of (6,5) and PFO in case of (7,5) SWCNTs) extending to the UV range.

## 2.3. Fitting PL spectra with a set of Voigt functions

PL spectra extracted from excitation-emission maps were fitted with a dedicated set of Voigt functions. For (6,5) SWCNTs, spectra were extracted for $\lambda_{ex} = 574$ nm and the initial positions of the fitted peaks were defined as 1130, 1160, 1220, 1340 and 1400 nm. Exemplary fits for (6,5) SWCNTs functionalized with BPO ($T = 100°C$, [R-BPO]/[CNT] = 42, $t = 3$h), 3-CN-BPO ($T = 70°C$, [R-BPO]/[CNT] = 11, $t = 3$h) and 3-OCH$_3$-BPO ($T = 100°C$, [R-BPO]/[CNT] = 168, $t = 3$h) are presented in Figure S4. Using this method, we calculated the exact spectral positions of the $E_{11}$ and $E_{11}*_{(1160)}$, as well as values of FWHM for $E_{11}$ and $E_{11}*$ peaks (Table S3). Such analysis enabled verification whether the $E_{11}*_{(1130)}$ feature emerged in the spectra.

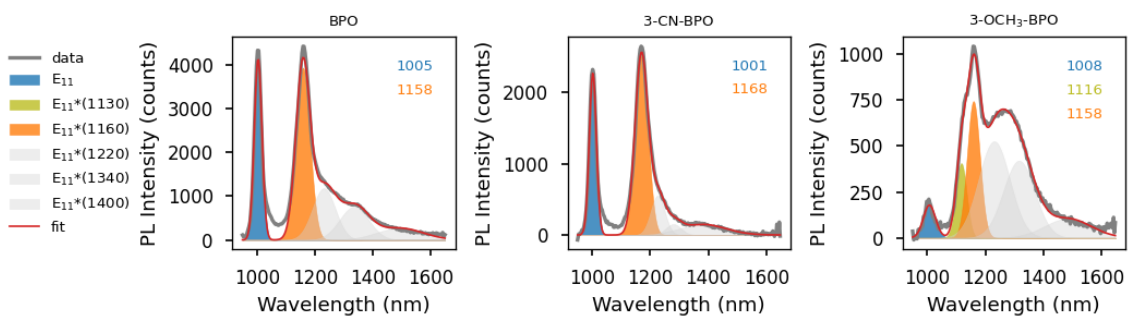

**Figure S4** Exemplary PL spectra of functionalized SWCNTs with visualization of the set of model Voigt peaks used in this work. Initial spectral positions of the defect peaks are shown in the legend, and the positions of fitted peaks are shown in plots as examples. The results obtained with this approach were used to calculate $E_{11}$ and $E_{11}*_{(1160)}$ peaks' positions and FWHM values (Table S3). As a result, it was possible to track the occurrence of the indicated peaks. The structure and positions of further peaks are not elucidated yet.

**2.4. PL spectra of functionalized (6,5) SWCNTs**

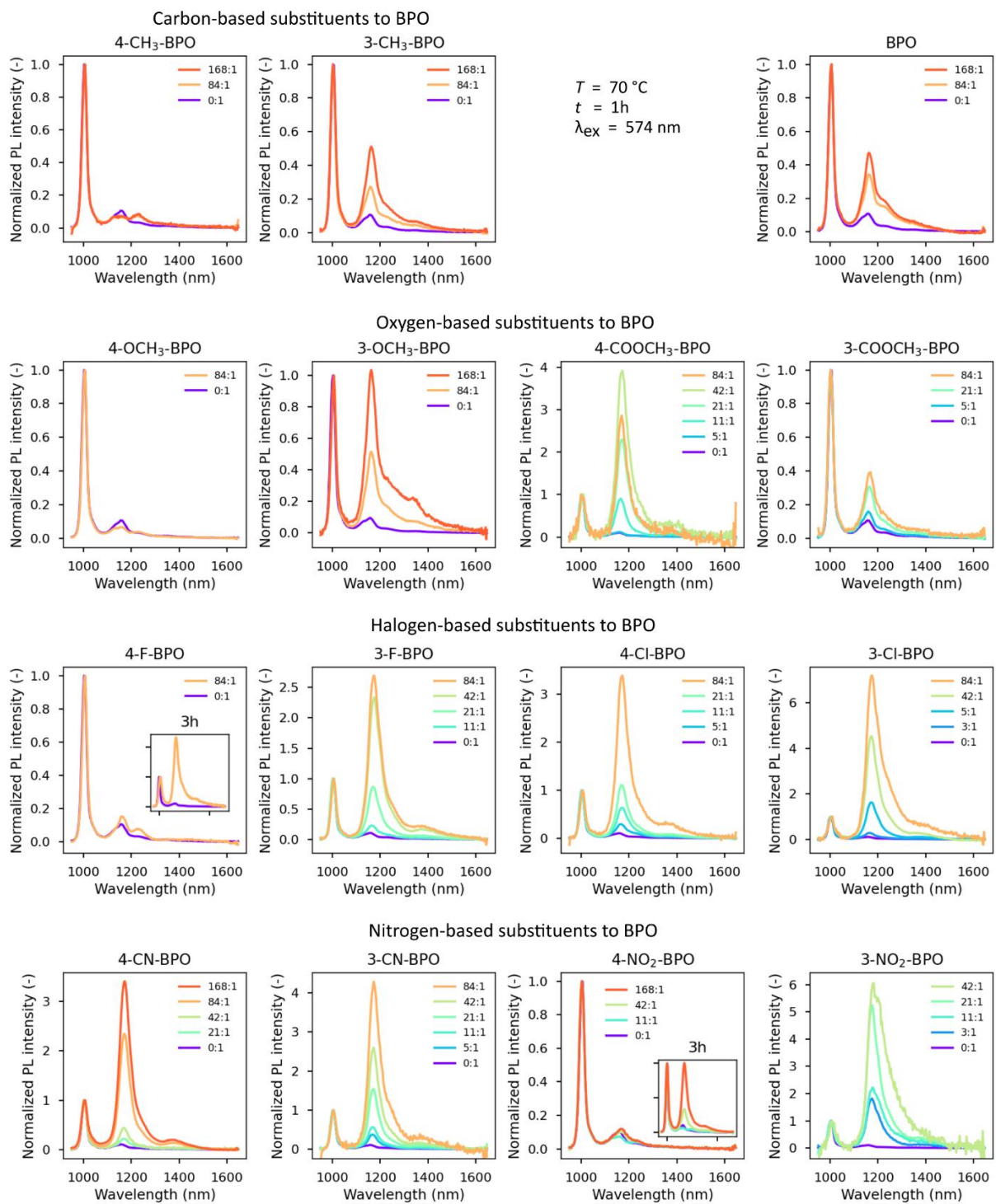

**Figure S5** PL spectra of polymer-extracted (6,5) SWCNTs functionalized using BPO derivatives in different molar ratios to SWCNTs. The SWCNTs were treated at 70 °C for 1 hour (except spectra for 4-F-BPO and 4-NO$_2$-BPO presented in the insets, which were obtained in 3 hours). Spectra were obtained for 574 nm excitation wavelength. For a visual comparison of defect densities, data were normalized to the maximum PL intensity of the E$_{11}$ peak.

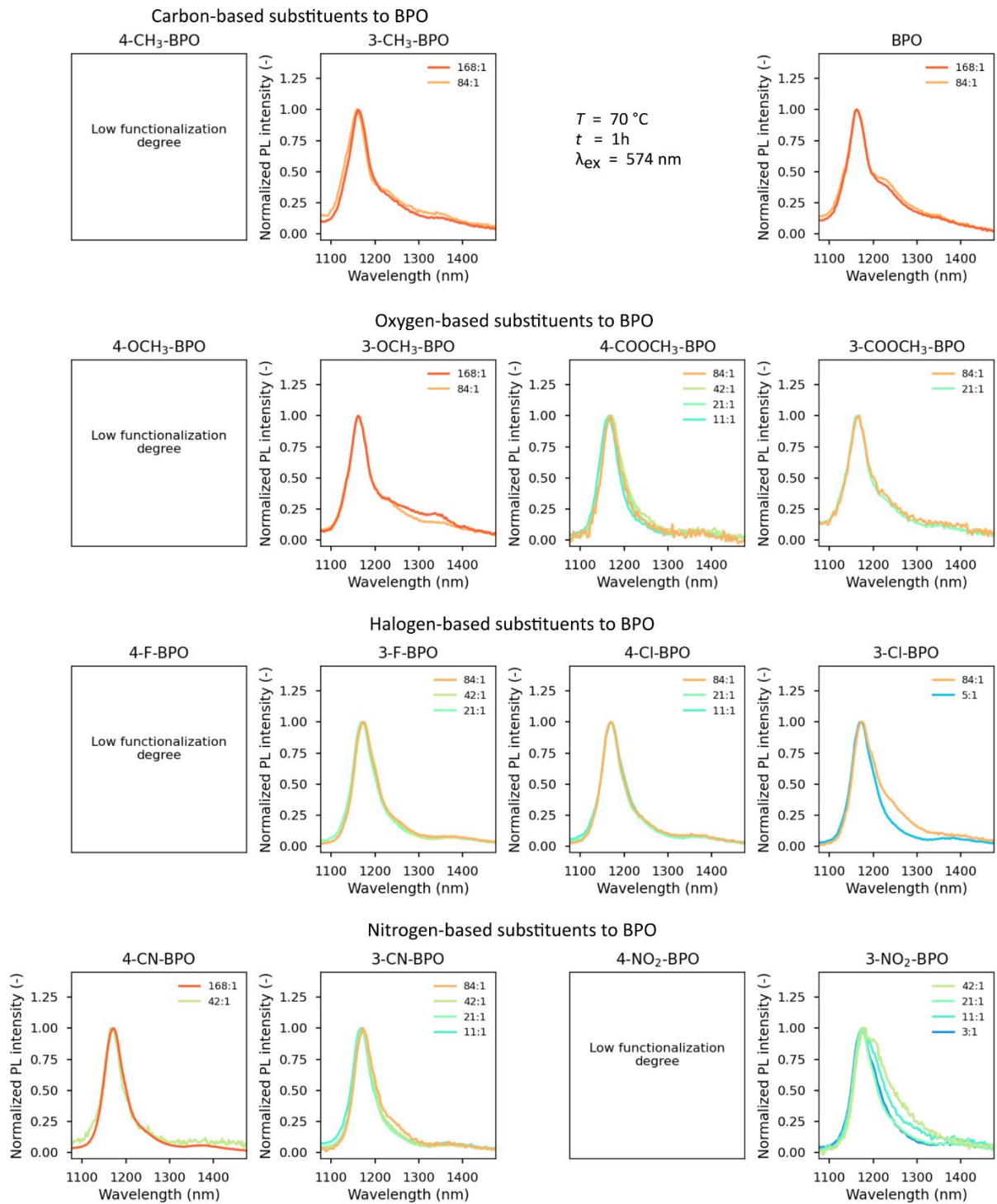

**Figure S6** PL spectra of polymer-extracted (6,5) SWCNTs functionalized using BPO derivatives in different molar ratio to SWCNTs. The SWCNTs were treated for 1 hour at 70 °C. Spectra were obtained for 574 nm excitation wavelength. For visual comparison of defect-originating peaks' shapes, data were normalized to the maximum PL intensity of $E_{11}*$ peak positioned at ca. 1160 nm and plotted from 1076 to 1476 nm. The indicated samples with low functionalization degree were not interpreted in this manner.

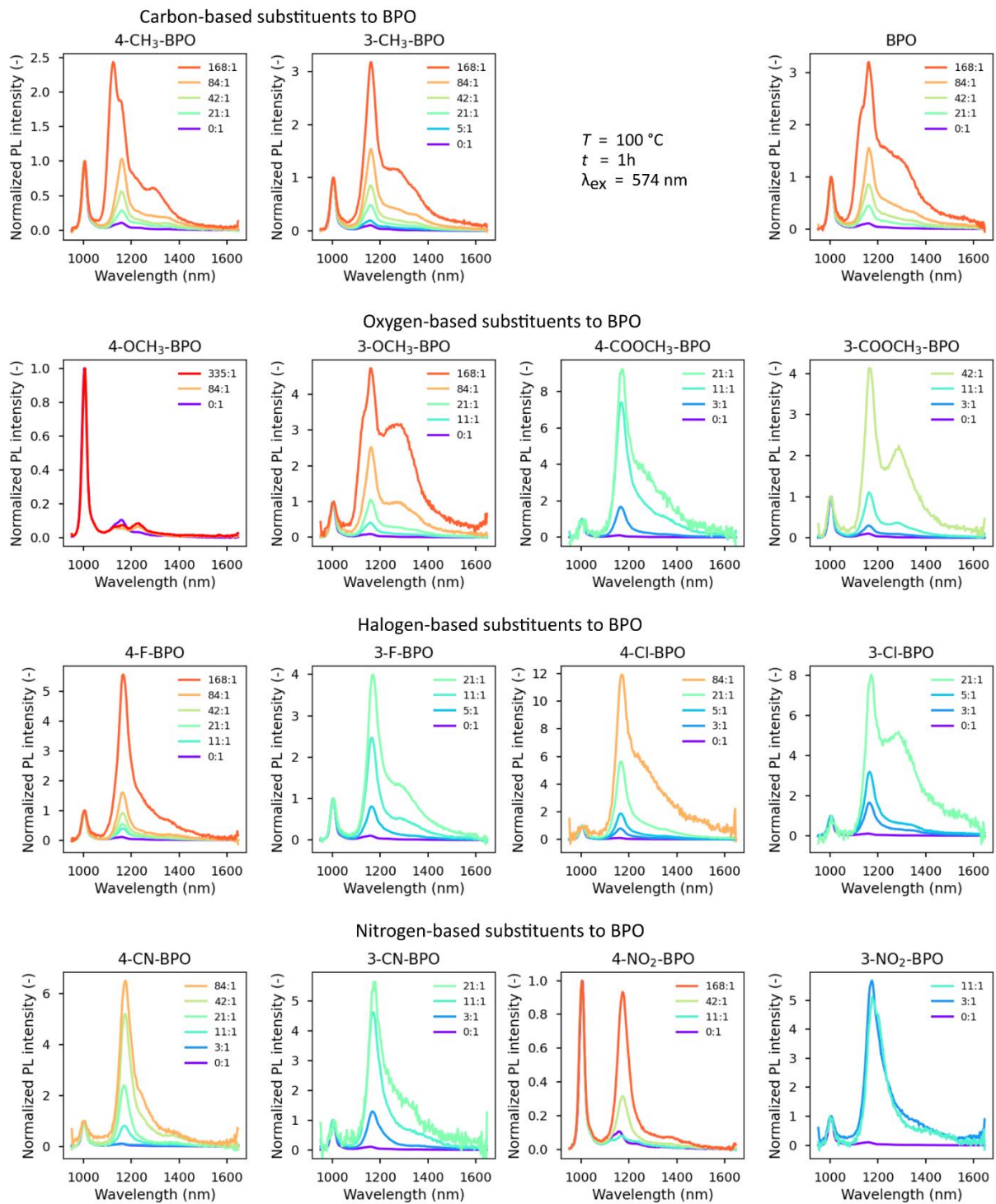

**Figure S7** PL spectra of polymer-extracted (6,5) SWCNTs functionalized using BPO derivatives in different molar ratios to SWCNTs. The SWCNTs were treated for 1 hour at 100 °C. Spectra were obtained for 574 nm excitation wavelength. For a visual comparison of defect densities, data were normalized to the maximum PL intensity of the $E_{11}$ peak.

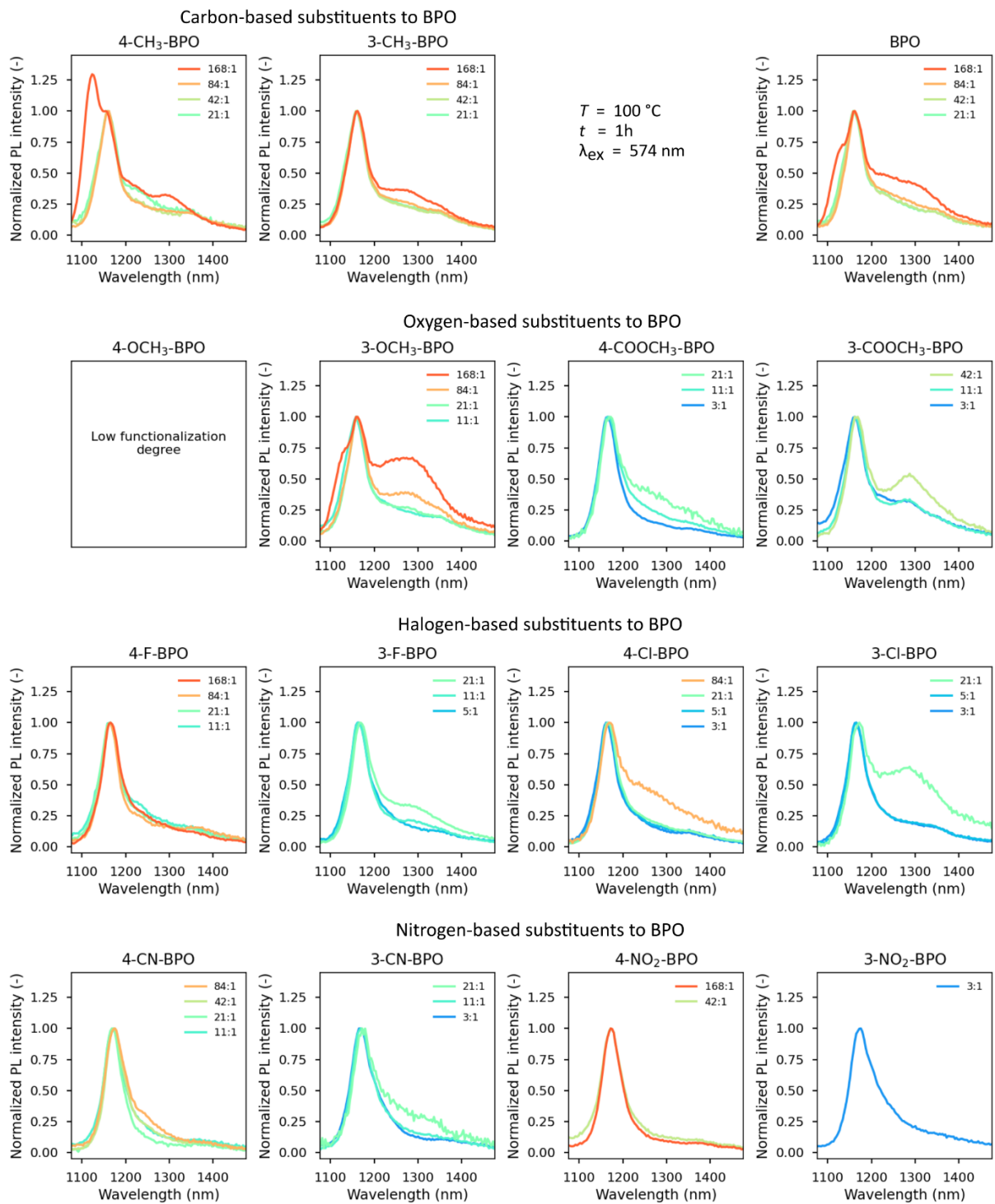

**Figure S8** PL spectra of polymer-extracted (6,5) SWCNTs functionalized using substituted BPO derivatives in different molar ratio to SWCNTs. The SWCNTs were treated at 100 °C for 1 hour. Spectra were obtained for 574 nm excitation wavelength. For visual comparison of defect-originating peaks' shapes, data were normalized to the maximum PL intensity of $E_{11}^*$ peak positioned at ca. 1160 nm and plotted from 1076 to 1476 nm. The indicated samples with low functionalization degree were not interpreted in this manner.

### 2.5. Properties of aryl peroxides

Atomic charges and electronic surface potentials for the benzoyloxyl radicals generated from the indicated BPO derivatives were calculated using Entos Envision, an open, online interactive platform for molecular simulation and visualization. This platform utilizes rapid semi-empirical calculations based on the GFN1-xTB method [12]. The functionalization of SWCNTs using various BPO derivatives revealed significant differences in reactivity depending on the substituents' electronic and steric properties (Table S2). Hence, the radical generation conditions and substituent effects in tailoring SWCNT optical properties determine the outcome of the SWCNT functionalization [13].

**Table S2** The properties of BPO derivatives: Hammett constant σ, atom-centered charges and electrostatic potentials.

| Substituent to BPO | Hammet const. σ (-) [14] | Atom-Centered Charges (-) | | | | Electronic Surface Potential (-) | | | |
|---|---|---|---|---|---|---|---|---|---|
| | | $C_{phenyl}$ | $C_{carbonyl}$ | $O_1$ | $O_2$ | $C_{phenyl}$ | $C_{carbonyl}$ | $O_1$ | $O_2$ |
| -H | 0 | 0.0674 | 0.528 | -0.446 | -0.387 | 0.0619 | -0.280 | -0.172 | 0.159 |
| 4-CH₃ | −0.170 | 0.0662 | 0.525 | -0.452 | -0.392 | 0.0588 | -0.286 | 0.169 | 0.169 |
| 3-CH₃ | −0.069 | 0.0678 | 0.526 | -0.453 | -0.392 | 0.0551 | -0.287 | 0.156 | 0.157 |
| 4-OCH₃ | −0.268 | 0.0459 | 0.551 | -0.41 | -0.409 | 0.0405 | -0.315 | 0.187 | 0.186 |
| 3-OCH₃ | 0.115 | 0.0662 | 0.524 | -0.464 | -0.398 | 0.0532 | -0.292 | 0.166 | 0.157 |
| 4-COOCH₃ | 0.450 | 0.0761 | 0.525 | -0.445 | -0.388 | 0.0654 | -0.276 | 0.175 | 0.161 |
| 3-COOCH₃ | 0.370 | 0.0682 | 0.526 | -0.445 | -0.387 | 0.0678 | -0.278 | 0.172 | 0.159 |
| 4-F | 0.062 | 0.0611 | 0.525 | -0.439 | -0.385 | 0.0734 | -0.274 | 0.175 | 0.161 |
| 3-F | 0.337 | 0.0733 | 0.526 | -0.441 | -0.385 | 0.0652 | -0.274 | 0.175 | 0.161 |
| 4-Cl | 0.227 | 0.0628 | 0.517 | -0.456 | -0.403 | 0.0623 | -0.287 | 0.169 | 0.155 |
| 3-Cl | 0.373 | 0.0672 | 0.519 | -0.459 | -0.403 | 0.0585 | -0.288 | 0.169 | 0.156 |
| 4-CN | 0.830 | 0.0747 | 0.527 | -0.438 | -0.382 | 0.0757 | -0.269 | 0.178 | 0.164 |
| 3-CN | 0.560 | 0.0700 | 0.527 | -0.439 | -0.383 | 0.0731 | -0.271 | 0.177 | 0.163 |
| 4-NO₂ | 0.778 | 0.0862 | 0.528 | -0.423 | -0.371 | 0.0898 | -0.248 | 0.189 | 0.174 |
| 3-NO₂ | 0.720 | 0.0739 | 0.528 | -0.422 | -0.373 | 0.0952 | -0.251 | 0.188 | 0.170 |

Unfortunately, establishing a single universal set of conditions (temperature, time, and concentration) for all investigated BPO derivatives is not possible. The surface modification of SWCNTs and radical chemistry are inherently complex. The attachment of various types of defects occurs at distinct reaction rates and depends on a broad cross-section of factors, including temperature, radical generation kinetics, reactivity of different radicals, recombination phenomena, or the presence of interfering small molecules (solvents, radical stabilizers, impurities, humidity, etc.). Nonetheless, the structure and electronic characteristics of BPO derivatives are critical to the reactivity of the resulting radicals, thereby governing their effectiveness as SWCNT functionalization agents. In particular, the positioning of the functional group relative to the peroxy moiety critically modulates the electron density distribution and steric accessibility at the reactive sites [15]. Additionally, the morphology of the resulting defects is highly dependent on both steric and electronic factors, making a predictive understanding of the behavior of individual radical sources essential [13,15,16]. For nearly all investigated radical BPO derivatives (except 4-OCH₃-BPO), suitable conditions and reaction mixture compositions can be identified to enable functionalization across a wide range of defect densities. Given the vast number of possible temperature and concentration combinations, the ideal parameters identified in this study are not exhaustive. For specific initiators, after analysis of the decomposition profile and kinetics, the prime conditions may fall within a narrower range between the investigated temperatures. Refining these parameters further could improve precision in controlling reaction kinetics.

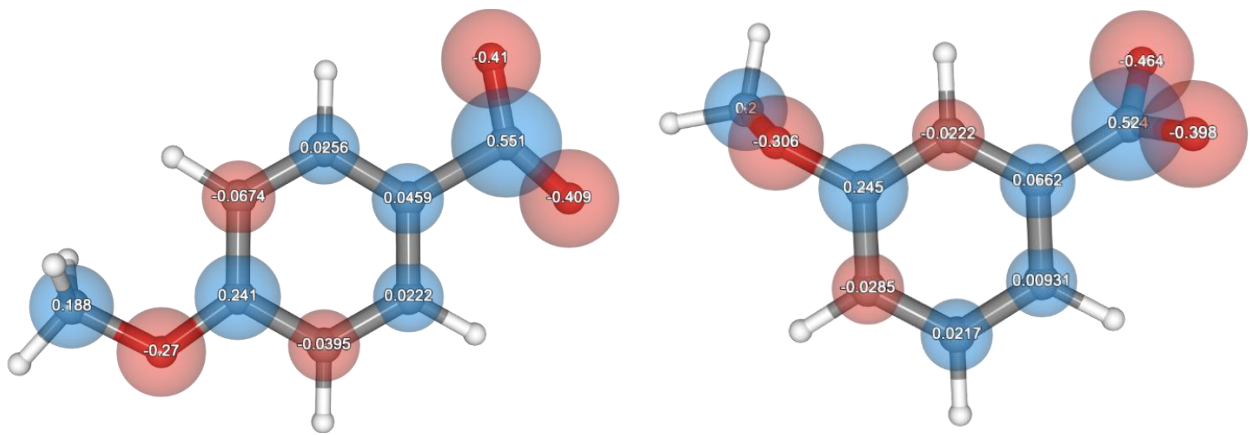

**Figure S9** Calculated Atom-Centered Charges for radicals derived from (left) 4-OCH₃-BPO and (right) 3-OCH₃-BPO using Entos Envision [12].

### 2.6. Kinetic control over functionalization process, defect stability and radical quenching

Given the radical nature of reactions initiated by BPO derivatives, it is expected that once started, this process will proceed until the generated radicals are depleted. Moreover, the presence of radicals in the reaction mixture can induce generation of secondary radicals, which can sustain the radical process [17]. On the other hand, our previous studies indicated that after 3 hours of heating BPO at 100°C, despite being a relatively unreactive initiator, it is already wholly decomposed. We decided to investigate whether the reaction stops after the process is completed and, if not, how this affects the optical properties of the material.

A partial answer to this question can be found in the already presented results, *i.e.*, in the plots of the $E_{11}*/E_{11}$ area ratio versus time for various temperatures and BPO-based reactants (Figure 3a,c). To a large extent, reactions carried out at 100°C for 3 hours reached spectral stabilization, meaning that further heating did not alter the PL emission. However, in case of higher concentrations of more stable peroxides, the reaction was not complete. Therefore, we decided to re-examine the SWCNTs functionalized using different BPO derivatives, after a long period, to verify whether the desired optical characteristics are preserved. A compilation of stability tests for a broad range of initiators and concentrations is presented in Figure S10 and Figure S11. Interestingly, in dozens of samples tested over periods ranging from 3 days to 8 months, we observed well preserved PL emission, close to identical to previously measured. Only in case of large BPO/SWCNT ratios, which led to high defect density in SWCNT structure, the samples became over-functionalized with time. When the reaction was conducted at 100 °C (Figure S10), and the BPO derivative concentration was appropriately low, stabilization of the spectrum was observed, despite continued heating or months-long storage (Figure S12a shows representative PL spectra), indicating the depletion of radicals in the reaction mixture. However, at 70 °C, it was not possible to eliminate the radicals within the given time, which caused the reactions to proceed further during storage for up to 4 months (Figures S11 and S12b).

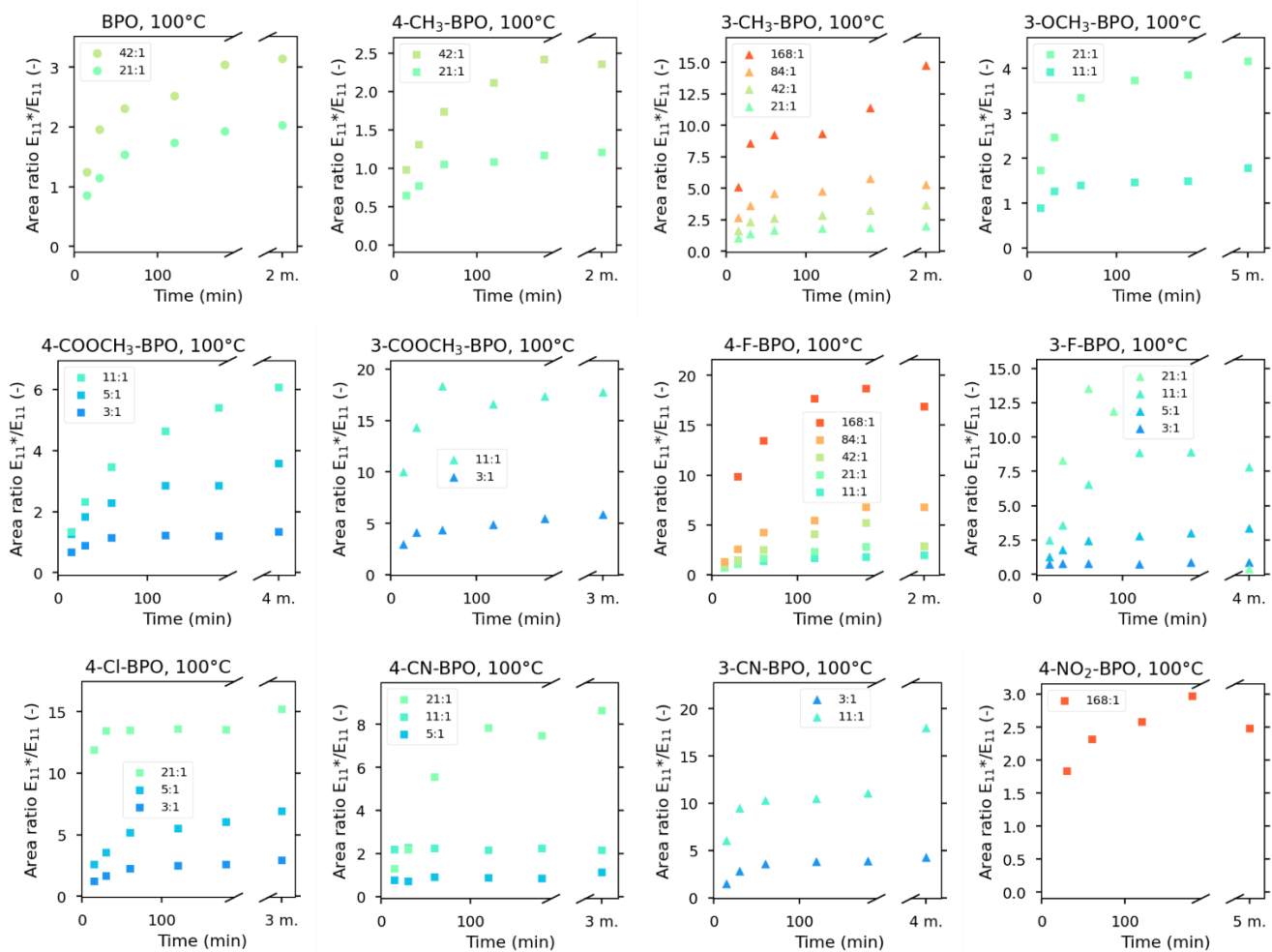

**Figure S10** Comparison of values of $E_{11}*/E_{11}$ area ratios obtained in reactions conducted at 100°C. (6,5) SWCNTs were reacted with radical initiator specified in the title, and its molar ratios with respect to SWCNTs in the legends. After the brake on horizontal axis, there are shown values obtained in measurements performed after long time (e.g. 3 m. means 3 months).

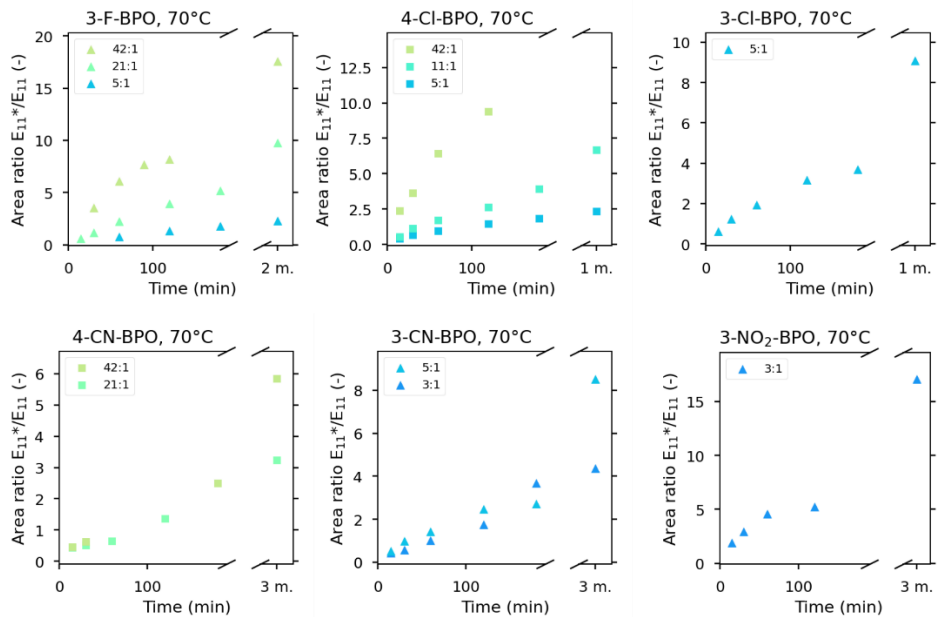

**Figure S11** Comparison of values of $E_{11}*/E_{11}$ area ratios obtained in reactions conducted at 70°C. (6,5) SWCNTs were reacted with radical initiator specified in the title, and its molar ratios with respect to SWCNTs in the legends. After the brake on horizontal axis there are shown values obtained in measurements performed after long time (e.g. 3 m. means 3 months).

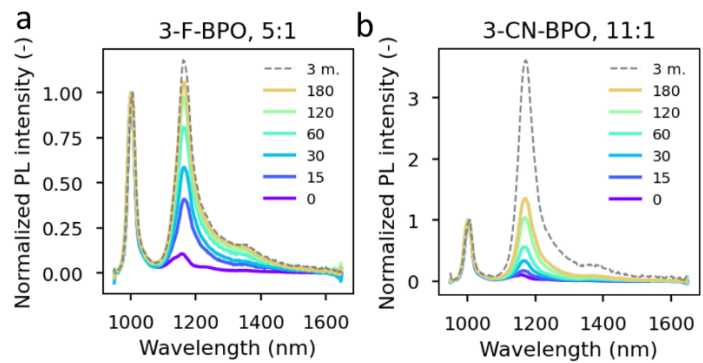

**Figure S12** Exemplary PL spectra of (6,5) SWCNTs reacted with 3-F-BPO and 3-CN-BPO at a) 100 and b) 70°C. Spectra obtained when samples were re-measured after 3 months of storage were plotted with dashed lines.

To slow down or halt the SWCNT functionalization process when exposed to high concentrations of the reactant, employing radical scavengers presents a viable approach. Radical scavengers are compounds that react with free radicals to terminate radical chain reactions, effectively quenching their reactivity. To this end, we investigated the use of radical scavengers such as TEMPO (2,2,6,6-Tetramethylpiperidine-1-oxyl), BHT (Butylated Hydroxytoluene), and PBN (N-tert-Butyl-α-phenylnitrone) to control and potentially freeze the functionalization reaction in situ [18]. We specifically chose SWCNTs functionalized at 70 °C for these experiments, as previous results indicated the persistent presence of active radicals at this temperature, capable of driving further reactions upon reheating or prolonged storage.

We first conducted preliminary tests to assess the impact of radical scavengers added at the beginning of the functionalization process. In the first experiments, TEMPO, BHT, or PBN were introduced into the reaction mixture at a two-fold molar excess relative to the initial benzoyloxy radical concentration. The aim was to verify if these scavengers could intercept the radicals, thus preventing the processing of undesired SWCNT functionalization. PL spectra were recorded after 60 minutes of heating at 70°C (Figure S13a, reference spectra without scavengers shown in purple). While not completely arresting the reaction, TEMPO exhibited promising behavior. It effectively limited the extent of functionalization, resulting in a substantially smaller increase in the $E_{11}*/E_{11}$ peak ratio, and, importantly, maintaining spectral stability and fluorescence intensity. This suggests that TEMPO successfully scavenges a significant portion of the generated radicals, thereby hindering further defect formation. In reaction mixtures containing BHT as a radical scavenger, we observed a significantly reduced increase in the defect-induced $E_{11}*/E_{11}$, indicating a suppression of SWCNT functionalization. PBN was the least effective and the spectrum obtained with the addition of this scavenger closely resembled the reference one.

To further investigate the effectiveness of radical scavengers, we conducted a subsequent experiment using radicals derived from 4-Cl-substituted BPO. After performing the functionalization reaction for one hour at 70°C, and recording the initial PL spectrum, we divided the sample into four aliquots: a control sample (no scavenger added), and the samples with TEMPO, BHT or PBN added (scavengers at a three-fold molar excess relative to the initial aryl peroxide concentration). PL spectra of these three samples were then measured after one month of storage at room temperature (Figure S13b). In the control sample without a scavenger, the reaction continued to progress during storage, as evidenced by a further increase in the defect density gauged by the $E_{11}*/E_{11}$ peak area ratio. In the sample treated with TEMPO, the defect density remained essentially unchanged clearly demonstrating effective inhibition of further functionalization. Both BHT and PBN failed to inhibit further functionalization. This observation further supports the notion that BHT or PBN may not act as radical scavengers in our system. In addition, they may potentially participate in complex side reactions and the present hydroxyl group can promote peroxide decomposition under these conditions.

Considering the collective observations from these scavenger experiments, the precise control over initiator concentration emerges as a critical factor for achieving optimal SWCNT functionalization. Careful adjustment of the initial BPO derivative concentration and the type of BPO derivative to match the desired defect density in SWCNTs appears to be a more effective and cleaner strategy than relying on radical scavengers (e.g. TEMPO) or prone-to-considerable-material-loss filtration often engaged to ensure termination of the SWCNT functionalization process.

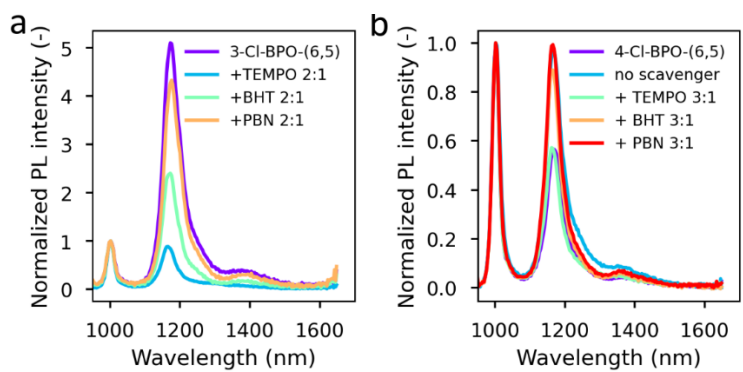

**Figure S13** Spectra obtained in reactions with radical scavengers: TEMPO, BHT and PBN. a) Selected scavenger in 2:1 molar ratio to radical initiator was present in the reaction mixture with SWCNTs during heating. b) Selected scavenger was added immediately after finished reaction (4-Cl-BPO-(6,5)) to stop the reaction; control reaction with no scavenger was also conducted. Spectra were remeasured after 1 week of storage at room temperature.

### 2.7. Density of defects measured by Raman spectroscopy

**Table S3** Maximum values of $E_{11}*/E_{11}$ area ratios and $I_D/I_G$ intensity ratios of in spectra of (6,5) SWCNTs thermally reacted with R-BPO in certain conditions: substituent R in BPO structure, reaction temperature (70 or 100°C) and molar excess of radical initiator over SWCNTs [R-BPO]/[CNT].

| | R | [R-BPO] /[CNT] | $E_{11}*/E_{11}$ | $I_D/I_G$ | [R-BPO] /[CNT] | $E_{11}*/E_{11}$ | $I_D/I_G$ | [R-BPO] /[CNT] | $E_{11}*/E_{11}$ | $I_D/I_G$ |
|---|---|---|---|---|---|---|---|---|---|---|
| | 4-CH₃ | 84 | 0.5 | 0.32 | 168 | 0.5 | 0.32 | - | - | - |
| | 4-Cl | 5 | 3.8 | 0.29 | 11 | 7.3 | 0.34 | 84 | 12.2 | 0.45 |
| 70 °C | 3-F | 11 | 2.7 | 0.15 | 21 | 10.0 | 0.38 | 84 | 9.5 | 0.32 |
| | 3-CN | 11 | 8.6 | 0.09 | 21 | 18.9 | 0.45 | - | - | - |
| | 4-CN | 84 | 11.0 | 0.14 | 168 | 11.5 | 0.28 | - | - | - |
| | 4-OCH₃ | 84 | 0.6 | 0.03 | 335 | 0.7 | 0.06 | - | - | - |
| | 4-CH₃ | 21 | 1.4 | 0.07 | 42 | 2.6 | 0.12 | 168 | 8.34 | 0.29 |
| 100 °C | -H | 21 | 2.1 | 0.06 | 168 | 5.9 | 0.31 | - | - | - |
| | 3-OCH₃ | 11 | 2.0 | 0.04 | 84 | 8.5 | 0.09 | 168 | 37.9 | 0.37 |
| | 3-COOCH₃ | 11 | 6.4 | 0.1 | 42 | 20.4 | 0.18 | - | - | - |

## 2.8. Dependence of PL redshift on Hammett substituent constant

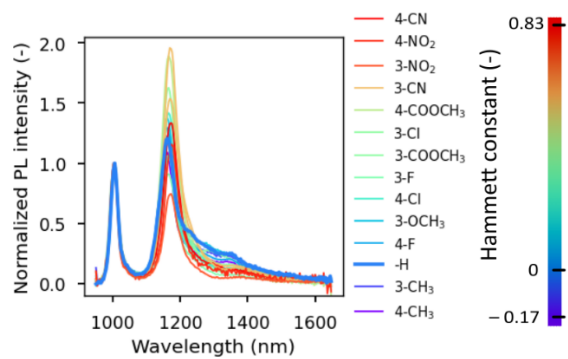

**Figure S14** PL spectra obtained for different derivatives of BPO (substituents are listed in the legend). Line colors were selected in linear dependence on the Hammet substituent constant, as the colorbar shows.

**Table S4** Spectral parameters of (6,5) SWCNTs functionalized using substituted BPOs as radical source. [R-BPO]/[CNT] – molar ratio of radical initiator to (6,5) SWCNTs; $E_{11}^*/E_{11}$ area ratio – ratio of area of the defect peaks in range 1080-1600 nm compared to $E_{11}$ peak area; $E_{11}^*$-$E_{11}$ – difference in spectral positions of the centers of peaks' located around 1160 and 1000 nm, respectively.

| Substituent | 70°C | | | | | | | 100°C | | | | | | |
|---|---|---|---|---|---|---|---|---|---|---|---|---|---|---|
| | [R-BPO]/[CNT] | $E_{11}^*/E_{11}$ area ratio | $E_{11}$ position (nm) | $E_{11}^*$ position (nm) | $E_{11}^*$-$E_{11}$ (nm) | $E_{11}$ FWHM (nm) | $E_{11}^*$ FWHM (nm) | [R-BPO]/[CNT] | $E_{11}^*/E_{11}$ area ratio | $E_{11}$ position (nm) | $E_{11}^*$ position (nm) | $E_{11}^*$-$E_{11}$ (nm) | $E_{11}$ FWHM (nm) | $E_{11}^*$ FWHM (nm) |
| -H | - | - | - | - | - | - | - | 42 | 4.6 | 1005 | 1161 | 156 | 30 | 57 |
| 4-CH₃ | - | - | - | - | - | - | - | 42 | 3.6 | 1003 | 1160 | 156 | 30 | 47 |
| 3-CH₃ | 168 | 5.5 | 1004 | 1162 | 158 | 32 | 58 | 42 | 4.3 | 1005 | 1161 | 156 | 29 | 56 |
| 4-OCH₃ | - | - | - | - | - | - | - | - | - | - | - | - | - | - |
| 3-OCH₃ | 168 | 4.0 | 1004 | 1162 | 158 | 29 | 58 | 21 | 4.9 | 1006 | 1161 | 155 | 29 | 57 |
| 4-COOCH₃ | 11 | 4.3 | 1004 | 1166 | 162 | 34 | 54 | 3 | 3.4 | 1002 | 1167 | 165 | 28 | 54 |
| 3-COOCH₃ | - | - | - | - | - | - | - | 5 | 4.1 | 1003 | 1163 | 160 | 29 | 57 |
| 4-F | - | - | - | - | - | - | - | 21 | 3.9 | 1005 | 1160 | 155 | 32 | 57 |
| 3-F | 5 | 2.6 | 1005 | 1165 | 160 | 30 | 54 | 5 | 3.8 | 1004 | 1164 | 160 | 29 | 56 |
| 4-Cl | 5 | 3.9 | 1004 | 1166 | 161 | 30 | 52 | 3 | 3.4 | 1004 | 1164 | 160 | 28 | 56 |
| 3-Cl | 3 | 4.0 | 1005 | 1165 | 159 | 32 | 52 | - | - | - | - | - | - | - |
| 4-CN | 21 | 3.7 | 1004 | 1171 | 167 | 30 | 53 | 11 | 2.5 | 1003 | 1171 | 169 | 30 | 57 |
| 3-CN | 3 | 4.9 | 1006 | 1171 | 165 | 31 | 52 | 3 | 4.8 | 1003 | 1169 | 166 | 30 | 57 |
| 4-NO₂ | - | - | - | - | - | - | - | 168 | 3.0 | 1003 | 1172 | 169 | 29 | 55 |
| 3-NO₂ | 3 | 3.5 | 1003 | 1172 | 169 | 29 | 53 | - | - | - | - | - | - | - |

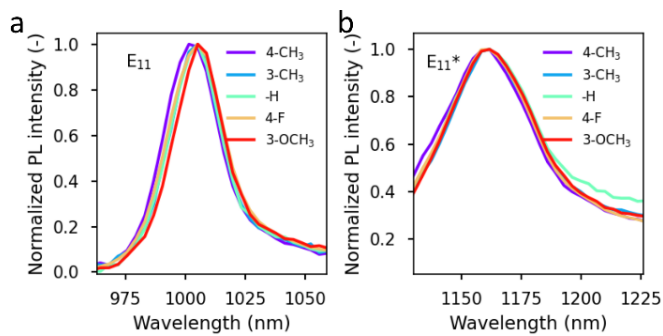

**Figure S15** Spectra obtained using radicals containing substituents of low Hammet constant values <0.2 resulted in very similar redshifts at 100 °C. The positions of the $E_{11}$ and $E_{11}^*$ peaks are shown in a) and b), respectively.

## 2.9. Functionalization of (7,5) SWCNTs

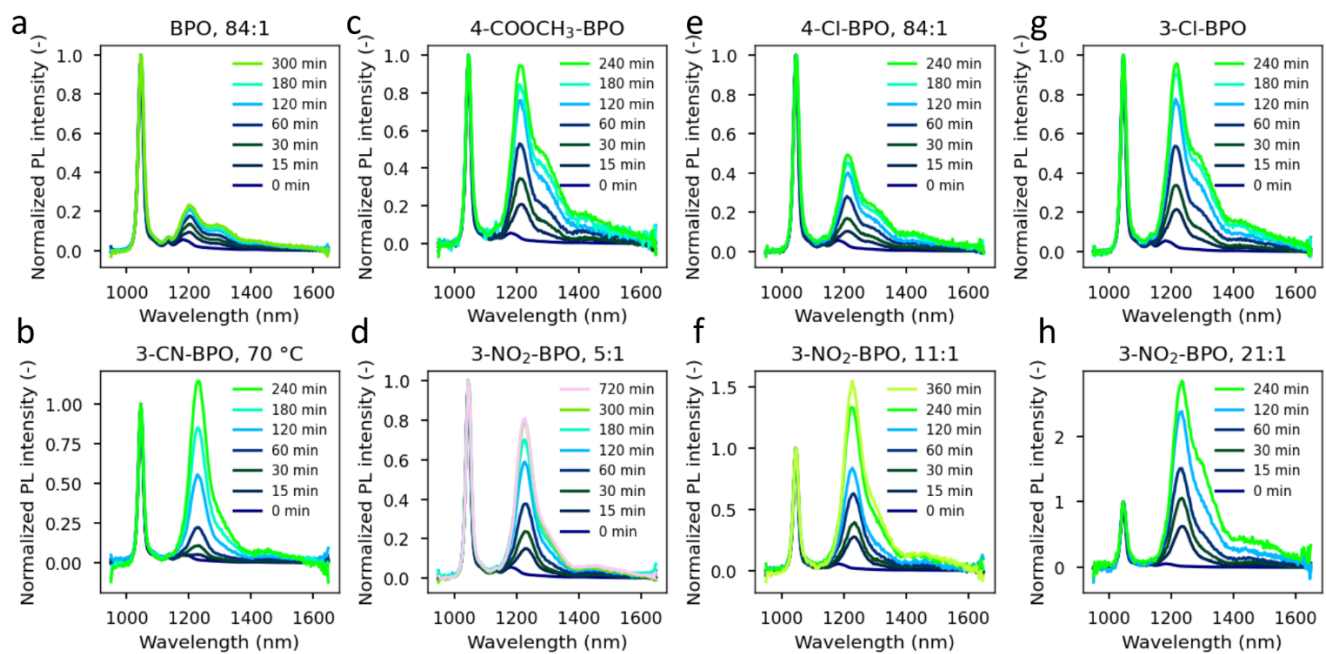

**Figure S16** PL spectra of (7,5) SWCNTs functionalized using BPO and its derivatives. Initiators are listed in the titles. The molar ratio of initiator to SWCNTs was 42:1 and the reaction temperature was 100°C, unless specified otherwise in the plot title. The spectra were obtained for 653 nm excitation wavelength.